\documentclass[aps,a4paper]{revtex4}
\usepackage{epsfig}
\usepackage{graphicx}
\usepackage{amsmath,amssymb,color}
\usepackage[english]{babel}

\parskip=\medskipamount

\newcommand{\eq}[1]{(\ref{#1})}
\newcommand{\fig}[1]{Fig.~\ref{#1}}

\newcommand{\be}{\begin{equation}}
\newcommand{\ee}{\end{equation}}
\newcommand{\ve}{\textbf}
\newcommand{\eps}{\varepsilon}

\makeindex

\begin{document}

\title{Dynamics of polymers: classic results and recent developments}

\author{Mikhail V. Tamm$^{1,2}$ and Kirill Polovnikov $^{1,3}$}


\address{ $^1$ Faculty of Physics, Moscow State University, 119991,
Moscow, Russia \\
$^2$ Department of Applied Mathematics, National Research University Higher School of Economics, 101000,
Moscow, Russia \\
$^3$ Skolkovo Institute of Science and Technology, 143005, Skolkovo, Russia \\
}

\begin{abstract}
In this chapter we review concepts and theories of polymer dynamics. We think of it as an introduction to the topic for scientists specializing in other subfields of statistical mechanics and condensed matter theory, so, for the readers reference, we start with a short review of the equilibrium static properties of polymer systems. Most attention is paid to the dynamics of unentangled polymer systems, where apart from classical Rouse and Zimm models we review some recent scaling and analytical generalizations. The dynamics of systems with entanglements is also briefly reviewed. Special attention is paid to the discussion of comparatively weakly understood topological states of polymer systems and possible approaches to the description of their dynamics.
\end{abstract}

\maketitle

\section *{Introduction}

Several last decades have seen great progress in our understanding of the physical principles underlying equilibrium structure, properties and dynamics of systems consisting of long polymer molecules. In 1930s-50s the simple basic principles of polymer physics were mostly developed by physical chemists, most importantly P. Flory, W. Kuhn, W. Stockmayer, P. Rouse, B. Zimm, V. Kargin, who were mostly motivated by possible applications in chemical technology. The second, glorious stage in the history of the field, lasting from 1960s to 1990s, was triggered by the inflow of talent from `big' theoretical physics, most notably P.-G. de Gennes, I.M. Lifshitz, S.F. Edwards, and their students. This second generation was motivated primarily by the desire to understand the origins and functioning of life (indeed, starting from the middle of 20th century it became crystal clear that polymer molecules play a crucial role in the functioning of all living beings). It was able to uncover deep connections between polymer physics and various other fields of condensed matter physics, such as field theory, statistical physics of disordered media, and, most notably, theory of second order phase transitions.

Since then, polymer physics has become an integral part not only of the soft matter field, but also of the more general statistical mechanics, to the point that some topics related to equilibrium properties of polymer chains (so-called self-avoiding random walks and polymer-magnetic analogy) are often covered in the curriculum of contemporary courses in statistical mechanics\cite{sethna,cardy}. On the other hand, a similarly rich and beautiful topic of polymer dynamics is, to the opinion of the authors, less widely known by the scientific community outside the narrow polymer physics field. This chapter aims to be a small contribution to the cause of overcoming this deficiency. 

The chapter consists of three large sections. The first section aims to give the reader a short introduction into the statistical theory of equilibrium polymer systems to the extent needed to understand the dynamics which will come later. First half of the material presented here is widely accepted matter which is covered in much more details in classical textbooks and monographs on polymer physics \cite{degennes,gr_khokhlov,rubinstein}. The second half of this section covers the properties of a class of polymer systems which is much less understood, namely, the systems whose equilibrium properties are governed by topological interactions of the chains. It has been known for a long time that, in particular, melts of nonconcatenated polymer rings, are an example of such a system whose equilibrium properties are very different from those of usual linear polymers. There is more and more evidence that such states can be observed in a variety of contexts: in particular, they seem to be a natural model of how chromosomes are packed in living cells. A proper theory of this `topological' state of polymer materials coming from first principles is still lacking (and, as many say, is the last big unsolved problem of polymer physics). However, many facts about equilibrium properties of such systems are well-established, and we review them briefly here.

The second section, which plays the central role in the chapter, discussed the dynamics of polymer systems to the extent when it is not influenced by the entanglements of chains with each other. Here we once again combine textbook material\cite{degennes,gr_khokhlov,rubinstein, doi_edwards} on classical Rouse\cite{rouse} and Zimm\cite{zimm} models of polymer dynamics with recent semi-analytical approaches developed recently in order to generalize this classical theories to the cases of systems with volume and topological interactions, and polymer chains in viscoelastic solvents.

 In the third section we discuss the limits of applicability of different theories, give reader a brief introduction to the reptation theory developed by de Gennes\cite{degennes_reptation} and Doi and Edwards\cite{doi_edw1,doi_edw2} in order to describe the dynamics of entangled linear polymers. We also briefly mention recent attempts to generalize reptation theory for the melts of ring polymers.

\section {Overview of the equilibrium properties of polymer chains, solutions, and melts}

\subsection{Ideal polymer chains}

Polymers are long sequences of chemically identical (or almost identical) monomer units. As a result of thermal fluctuations in the length of the chemical bonds connecting the monomer units and the angles between them, polymer chains are flexible. There is a characteristic chain length associated with this flexibility, which is usually called {\it persistence length} of the chain $l$. It is known that large-scale properties of long polymer chains (and of any chain fragments of length much larger than $l$) are largely universal and insensitive to a particular flexibility mechanism. 
Therefore, it is desirable to make the chain flexibility model as simple as possible. The most common examples are the model of a uniformly flexible chain and the beads-and-springs model. 

In the former model, the chain conformation is represented by a function $\ve r (s)$ where $s \in [0, L]$ is a linear coordinate along the chain, and $\ve r (s)$ is the position of the corresponding point in space. The energy to be associated with the chain flexibility is
\be
U_{id} \left( \{ \ve r(s) \} \right) = \int_0^L \kappa \left( \frac{\partial \ve r}{\partial s}\right)^2 ds,
\label{persistent}
\ee
where $\kappa$ is a flexibility parameter, $\kappa \sim T/l$ and $T$ is the temperature (which is hereafter measured in units of energy, so that $k_B \equiv 1$) . If chain flexibility is the only contribution to the energy of the system (the case of an {\it ideal polymer chain}), the corresponding partition function can be written down as
\be
Z_{id} = \int \exp ( - U_{id} \left( \{ \ve r(s) \} \right) /T) {\cal D} \ve r(s).
\label{persistent_partition}
\ee
Clearly, this representation bears a striking similarity to statistical weights of Brownian trajectories. And, in full analogy with the theory of Brownian motion, if one considers a long enough polymer chain starting from the origin ($\ve r(0)=\ve 0$), the position of the other end $\ve r(L)$ will be normally distributed with zero mean and dispersion proportional to $\kappa L/T$. 

The notion that ideal polymer chains are Gaussian at large enough lengthscales gives rise to the second model mentioned above: a simple discrete {\it beads-and-springs model}. Let us track, instead of the whole function $\ve r(s)$, just its values at equidistant points $\ve r_0 =\ve r(0), \ve r_1 = \ve r(L/N), \ve r_2 = \ve r(2L/N), ... ,\ve r_N= \ve r (L)$. If $L/N$ is sufficiently larger than the persistent length $l$ one can assume that parts of the chain between the tracked points are already Gaussian, and write the partition function of the whole chain in the following form:
\be
Z_{id} = \prod_{i=1}^N g\left(\ve r_{i-1} - \ve r_i \right); \; \; g(\ve r) = (2\pi a^2/ d)^{-d/2} \exp \left( -\frac{d \ve r^2}{2 a^2} \right) ,
\label{part_ideal}
\ee 
where $d$ is the space dimensionality. Here we introduce a new variable $a$ which is of order $\sqrt{\kappa L/ N T}$ and has a meaning of a mean-square distance between points (particles, or {\it beads}), which we are tracking. It is easy to see that the partition function \eq{part_ideal} is exactly the same as for a system of beads connected by ideal springs with elastic energy
\be
U(\ve r_0, ..., \ve r_N) = \frac{d T}{2a^2}\sum_{i = 1}^{N} \left(\ve r_{i-1} - \ve r_{i}\right)^2,
\label{energy_ideal}
\ee
which explains the name of the model. Importantly, the typical spatial size of an {\it ideal polymer coil} with the partition function \eq{part_ideal} is of order 
\be
R_{id} \equiv \sqrt {\langle (\ve r_N - \ve r_0)^2 \rangle}\sim \sqrt{N} a,
\label{r_ideal}
\ee
therefore the coil is a very diluted object: indeed, $N$ beads are dispersed in the volume of order $R^3$ resulting in a number density
\be
c \sim N R_{id}^{-3} \sim N^{-1/2} a^{-3}.
\label{density}
\ee

Let us emphasize once again that for long chains and for $N \gg 1$ the results of the continuous and discrete approaches based on energies \eq{persistent} and \eq{energy_ideal} are indistinguishable. Below we will rely on the discrete model as it is often easier to speculate about, but will divert to the continuous one whenever it will simplify the formulae.

\subsection{Volume interactions. Equilibrium states of a single chain}

The models introduced above describe the connectivity of a linear polymer chain. On top of that, monomer units of the chain interact by what is known as {\it volume interactions}, i.e. physical interactions of the units depending on their relative spatial position regardless of whether they are adjacent along the chain or not. Indeed, ideal polymer chain is flexible and there is a non-negligible probability that monomer units which are far from each other along the chain can find themselves close to each other in space and therefore interact by, for example, van-der-Vaals forces. Within the beads-and-springs model it is easy to modify the partition function of a single chain to take these interactions into account:
\be
Z = Z_{id} \exp \left( - U_{vol}(\ve r_0, ..., \ve r_N)/T \right); \; \; U_{vol}(\ve r_0, ..., \ve r_N) = \sum_{i<j} V\left( \ve r_i - \ve r_j \right), 
\label{part_volume}
\ee
and $V (\ve r)$ is the potential of volume interactions between two monomer units. 
Depending on the interplay between the interaction potential $V (\ve r)$ and the value of temperature $T$, volume interactions might lead to either swelling or contraction of the polymer coil compared to its ideal size $R_{id}$. Importantly, properties of these two limiting states -- the so-called {\it swollen polymer coil} ($R>R_{id}$) and {\it equilibrium polymer globule}($R<R_{id}$) -- are largely independent of the particular form of $V (\ve r)$, provided that it is short-ranged. More precisely, the equilibrium state of a polymer chain is controlled by the sign of the {\it second virial coefficient} of volume interactions:
\be
B(T) = \frac{1}{2} \int \left(1 - \exp(-V(\ve r)/T))\right) d \ve r.
\ee
If $B(T) > 0$, the dominant form of volume interaction is hard-core repulsion, and polymer is in a swollen state (``good solvent'' regime), if $ B(T) < 0$, attraction dominates and polymer is collapsed into a globule (``poor solvent'' regime). In the vicinity of the so-called $\Theta$-temperature, at which $B(\Theta)=0$, there is a narrow transition region of width $\Delta T \sim \Theta N^{-1/2}$ between these two states. Within this transition region one may, in the first approximation, think of polymer chains as being in almost ideal state. 

\begin{figure}[!t] \centering
\includegraphics[width=\textwidth]{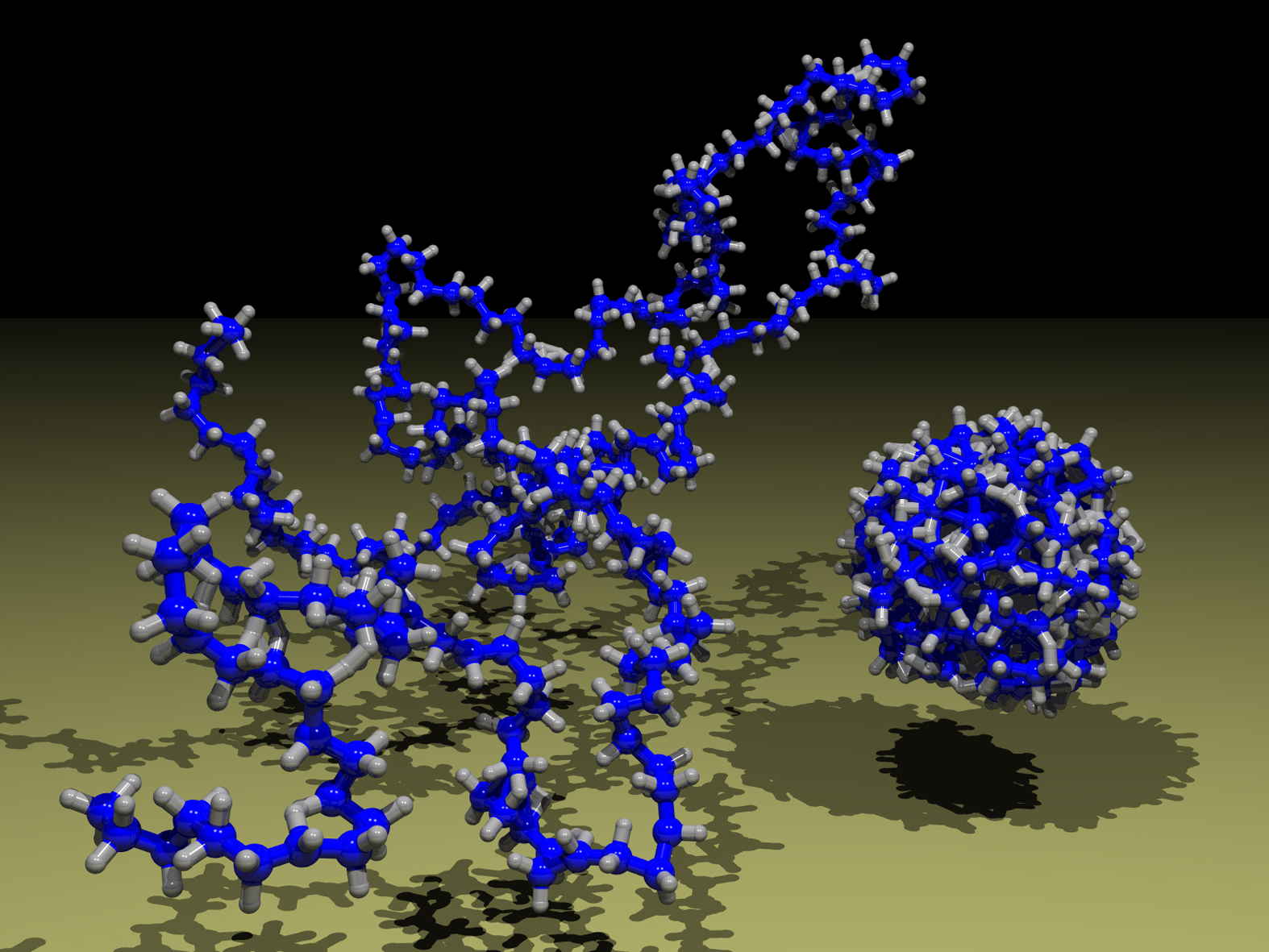}
\caption{Illustration of the different states of polymer chains. Ideal and swollen (left) polymer coils are extended states where conformations of the chains are statistically similar to trajectories of simple and self-avoiding, respectively, random walks. Equilibrium polymer globule (right) is a state where polymer chain forms a liquid-like droplet with positive surface tension between polymer and surrounding solvent. The picture is courtesy of P.G. Khalatur.}
\label{Khalatur}
\end{figure}

\subsection{Self-avoiding walks}

As mentioned above, the swollen state of a polymer coil is universal in a sense that it is independent of the nature of volume interactions and the form of their potential, provided that these interactions are short-range and predominantly repulsive (which corresponds to $B>0$). The simplest way of thinking of such interactions is to imagine a chain with {\it excluded volume} that prevents it from intersecting itself. In terms of the beads-and springs model, one can think of a hard-core repulsive potential: 
\be
V(\ve r) = \left\{ 
\begin{array}{ll}
\infty \; & \text{if } |\ve r| < r_0 \medskip \\
0,\; & \text{if } |\ve r| > r_0.
\end{array}
\right.
\ee
The resulting model is equivalent to a {\it self-avoiding random walk}. The corresponding partition function \eq{part_volume} of a chain with 0-th bead fixed at the origin and thr mean-square size of the chain $R = \sqrt {\langle (\ve r_N - \ve r_0)^2 \rangle} $ scale, respectively, as
\be
Z \sim z_0^N N^{-\gamma};\;\; R \sim N^{\nu}
\label{critical}
\ee 
with critical exponents $\gamma$, $\nu$ which are different from the values for ideal chains ($\gamma = 0$, $\nu = 1/2$) and depend on the dimensionality of space. Importantly, the conformation of a swollen polymer coil is fractal in a sense that large enough parts of the coil (of contour length $s$) demonstrate similar behavior $R(s) \sim s^{\nu}$ with the same exponent $\nu$.

The great breakthrough in understanding statistics of swollen polymer chains happened in the early 1970s when P.-G. de Gennes discovered \cite{deGennes_theorem}, by direct comparison of the corresponding diagrammatic expansions, an exact mapping of the self-avoiding walk problem onto the $n\to 0$ limit of the celebrated $O(n)$ model of magnetic phase transitions. As a result, it became possible to apply all the machinery of renormalization group, $\epsilon$-expansion, etc. for the calculation of the critical exponents $\gamma, \nu$ defined above. 

Two decades earlier P. Flory suggested\cite{flory,flory_book} a hand-waving quasi-mean-field computation of the exponent $\nu$ leading to 
\be
\nu = \left \{
\begin{array}{ll}
\frac{d}{d+2} & \text{ for } d \leq 4 \medskip \\
\frac{1}{2} & \text{ for } d > 4
\end{array}
\right.
\label{flory}
\ee
Impressively, this result is exact for all $d$ except $d=3$, where it overestimates the true value of $\nu$ by some 2\%, and $d=4$, where it is precise up to logarithmic corrections. As a result, polymer physicists often colloquially use \eq{flory} as an estimate for $\nu$ at all $d$. 

\subsection{Flory theorem. Polymer melts. Equilibrium globule state}

Before discussing the third equilibrium state of linear polymer chains - the {\it equilibrium globule}, we should say a few words about polymer melts, i.e. condensed phases consisting of many linear polymer chains with the same length and chemical structure. Will polymer chains in such a system be swollen or collapsed?

A somewhat counterintuitive answer to this question is ``neither''. Indeed, regardless of the nature of short-range interactions between the chains, in a melt they preserve ideal conformations on a large enough scale. This statement is known as {\it Flory theorem} in the literature. Qualitatively this fact can be understood as follows. The most entropically favorable state of a polymer chain is the state of an ideal coil. If, on top of the entropic considerations, there is an interaction energy, the chain might change its configuration to optimize the combination of energy and entropy: it may swell to reduce the number of monomer-to-monomer interactions if they are energetically unfavorable or collapse to increase their number if they are favorable. Now, polymer melt is a condensed state with essentially no voids. As a result, every monomer unit of the chain is always surrounded by the same number of neighbors, either belonging to the same chain or to one of the surrounding chains. If the monomer units of the chain under consideration and the surrounding chains are chemically equivalent, then the interaction energy is the same for monomers belonging to a single chain and to different chains. Therefore, interaction energy does not depend on chain conformations and, unable to optimize internal energy in any way, individual chains will take the most entropically advantageous ideal conformations. 

An equilibrium polymer globule formed by a collapsed linear chain can be thought about as a droplet of polymer melt (or concentrated polymer solution). If we consider a small volume of the globule far away from its surface, we will see many different chain fragments going through this volume, each of them having a locally ideal conformation. Meanwhile, on the surface of the globule there is a thin boundary region where polymer chains are substantially non-ideal: within this region they are ``reflected'' from the surface and go back into the bulk of the globule. In other words, conformation of a polymer chain in a globule state can be approximated by trajectory of a Brownian particle inside a spherical domain with reflecting boundary conditions. 

Importantly, in contrast to the ideal and swollen polymer conformations, the chain conformation in the equilibrium globule is not fractal: short chain fragments that fit between two reflections from the surface obey ideal chain statistics \eq{part_ideal} and their linear size scales as $R(s) \sim s^{1/2}$, while the linear size of long chain fragment is limited by the size of the globule and does not dependent on $s$:
\be
R(s) \sim \left \{
\begin{array}{ll}
s^{1/2} &\text{ for } s<s^*\sim N^{2/3}, \medskip \\
R(N) \sim N^{1/3} &\text{ for } s>s^*\sim N^{2/3},
\end{array}
\right. 
\label{radius_globule}
\ee 
where the size of the whole globule $R(N)$ follows immediately from the fact that it is a dense liquid-like object and the estimate $s^*\sim N^{2/3}$ is obtained by sewing the two regimes together.

\subsection{Semi-dilute solutions} 

In conclusion of this overview of the classical results of the equilibrium polymer theory, let us say a few words about polymer solutions. As mentioned above, the equilibrium density of an ideal polymer coil goes to zero as the length of the chain grows (see \eq{density}). This is true even more for the swollen polymer coils. As a result, polymer coils in a good solvent start to overlap at very small concentrations of the polymer. Therefore, two different concentration regimes of a polymer solution should be distinguished. In the {\it dilute} regime polymer concentration is so small that different polymer coils are not overlapping and are separated from each other by a volume of pure solvent. In the first approximation one can think that polymer coils in the dilute solution are independent systems with almost no interactions between them. In a {\it semi-dilute} solution concentration of the polymer is still small in a sense that the great majority of the space is taken by solvent: the {\it volume fraction} of the polymer $\phi = v c$, where $v$ is the excluded volume of the bead, is much less then unity. However, it is high enough to make different coils overlap and interact. The crossover between these two regimes happens at the {\it critical overlap concentration} which is equal to an equilibrium concentration inside a single isolated coil in the abundance of solvent.

Interestingly, in a semi-dilute polymer solution in good solvent the typical span covered by a single polymer chain reduces with the increase of concentration, the phenomenon caused by {\it partial screening of volume interactions}. Indeed, in the high-dilution limit single isolated chains cover a span of order $R(N) \sim N^{3/5}$ (see section 1.3), while the limit of high concentrations $\phi \to 1$ corresponds to polymer melts with $R(N) \sim N^{1/2}$ according to Flory theorem (see section 1.4). At intermediate concentrations $\phi$ there exist a critical length scale $\xi(\phi)$ which separates two different conformation regimes. This length has a meaning of the ``effective mesh size'' of the entangled solution: a ball of radius $\xi$ is on average intercrossed by exactly one polymer chain. Fragments of the chain shorter than $\xi$ are swollen and behave as if they lived in a dilute solvent: indeed, at this length scale the chance to encounter a foreign chain is small, and the chain fragment ``does not know'' that it belongs to a solution of many chains. On the scale larger than $\xi$ the chains interact significantly and Flory theorem holds. As a result, the following behavior is observed:
\be
R(s) \sim \left \{
\begin{array}{ll}
s^{3/5} &\text{ for } s< \xi ^{5/3} = \phi^{-5/4} \medskip \\
s^{1/2} &\text{ for } s> \xi ^{5/3} = \phi^{-5/4}
\end{array}
\right. ,
\label{radius_semidilute}
\ee 
where the connection between $\xi$ and $\phi$ is established by the aforementioned fact that on average there is exactly one chain fragment in a ball of radius $\xi$. The lengthscale $\xi$ is known in the literature as a size of {\it concentration blob}\cite{degennes}.

\subsection {Topologically regulated conformations of polymer chains}

The results briefly outlined in the previous five subsections have been well-established since as early as 1970s and constitute the main dogma of equilibrium polymer physics. However, it turns out that on top of the three main ``classical'' conformation states of polymer chains there exists another seemingly wide-spread conformational state, which we will call here {\it topological} or {\it topologically regulated}, and which is not yet completely understood. 

\begin{figure}[!t] \centering
\includegraphics[width=\textwidth]{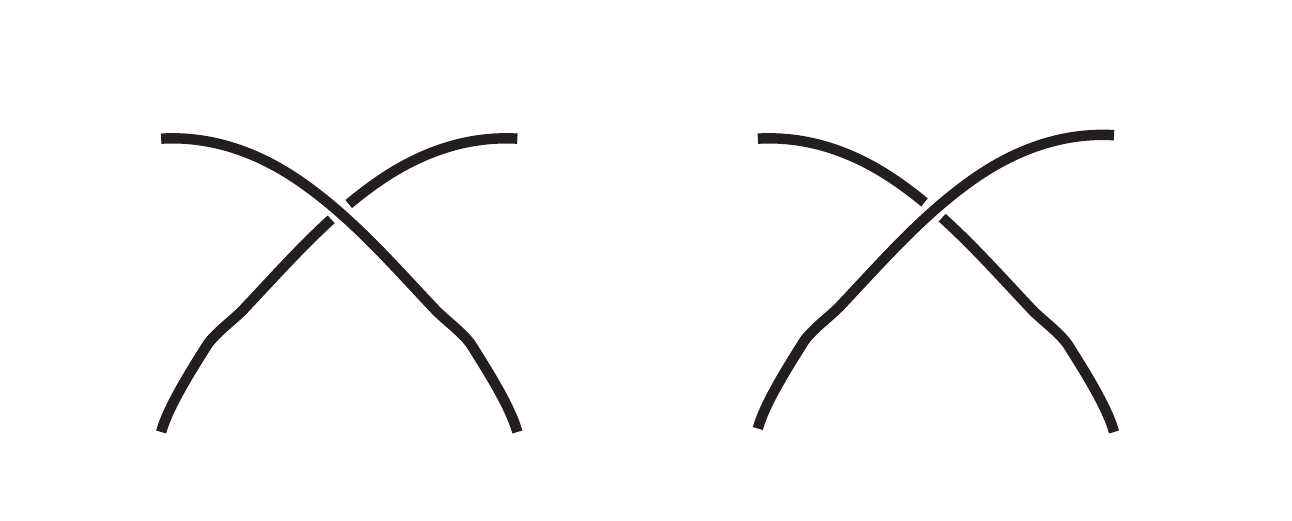}
\caption{Illustration of chain nonphantomness: conformations shown in the left and in the right are very close in phase space, a very small perturbation of coordinates can turn one into another. Nevertheless, since polymer chains are linear impenetrable objects such small perturbations are forbidden.}
\label{fig:1}
\end{figure}

First of all, let us note that there is a rather obvious property of polymer chains which we have not yet mentioned: polymer chains cannot go through each other, and therefore one cannot easily reach from conformation in the left of \fig{fig:1} to the conformation on the right. This property, called {\it non-phantomness of a chain}, is, generally speaking, absent in the beads-and-springs model. While we are talking about equilibrium properties of linear chains, one can argue that non-phantomness is irrelevant: indeed, it does not change which polymer conformations are allowed (and with what energy), it only changes which states are close to each other in the phase space. Therefore, if one waits long enough for the system to explore its whole phase space (which is pretty much a definition of equilibrium), the resulting statistical weights of conformations will be identical for phantom and non-phantom chains; only dynamics will differ. 

\subsubsection{ Polymer in the array of obstacles}

\begin{figure}[!t] \centering
\includegraphics[width=\textwidth]{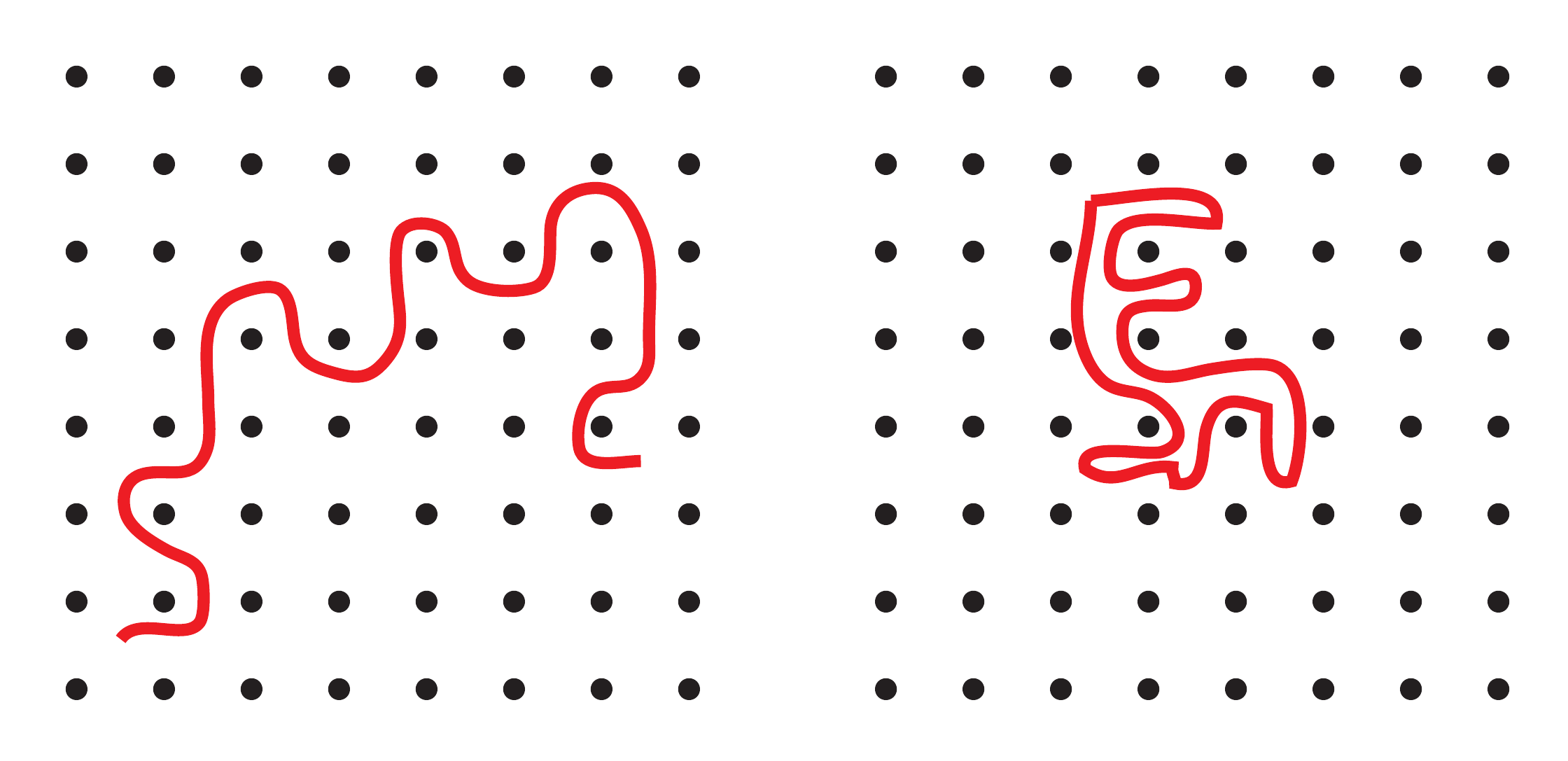}
\caption{Linear polymer chain (left) and polymer ring (right) in an array of impenetrable surfaces. The chains have similar length, note the difference in their spatial span. Note also that ring conformation can be described as a double-folded chain forming a randomly-branched structure.}
\label{fig:2}
\end{figure}

The situation changes dramatically as soon as a chain is no longer linear but is closed in a cycle (ring). Indeed, conformations of a non-phantom ring have an additional invariant of topological nature, which is the sort of a knot formed by the circular chain. If, for example, a chain is prepared in the unknotted state, it will stay in that state forever, thus all knotted conformations accessible to a phantom chain are forbidden for an unknotted non-phantom one. This fact leads to a radical difference between conformations of linear chains which can explore the whole phase space and those of circular ring chains which are confined to the part of phase space defined by a given value of the topological invariant. To get some insight into how that happens consider an example of a {\it polymer in array of obstacles} first considered in a seminal paper by Khokhlov and Nechaev more than 30 years ago\cite{khokhlov85}  (see \fig{fig:2}).

Consider a polymer on a plane with an additional constraint: a set of obstacles on the plane which the polymer chain cannot cross. Assume for simplicity that these obstacles are positioned in the nodes of a square lattice as shown in \fig{fig:2}. In equilibrium, conformations of a linear polymer chain in such an environment will have exactly the same large-scale statistics as in the unconstrained case: it will be defined by \eq{r_ideal} for ideal chains and \eq{critical}, \eq{flory} for swollen chains. The only effect of the obstacles will be to slow down the polymer dynamics. The situation becomes very different if the polymer chain is a ring. If such a chain is initially prepared in a way that there are no obstacles inside it, then the only conformations accessible to this chain will be of the type shown in \fig{fig:2}b; in these conformations the chain folds into a double line forming a randomly branched structure. Obviously, typical conformations of linear chains (\fig{fig:2}left) and rings (\fig{fig:2}right) are drastically different. They also have very different statistics: the cycle conformations can be mapped onto those of the {\it annealed randomly branched polymers} whose gyration radius is known to grow proportional to the number of monomers to the power 1/4 in the ideal case\cite{zs49,daoud_joanny}, and to the power 1/2 in the case of self-avoiding trees in 3d \cite{parisi_sourlas}. Interestingly, the number of conformations of such a ring with one point fixed can be effectively reduced to the problem of random walks on a tree or on a Lobachevsky-Riemann space\cite{khokhlov85,nechaev88}.

Note that conformations of ring polymers cannot be characterized by end-to-end distance as discussed above for linear chains. However, one can still speak of the gyration radius of the chain
\be
S^2 (N) = \sum_{i,j} \langle (\ve r_i - \ve r_j)^2\rangle.
\ee
For linear chains gyration radius always scales with the same exponents as the end-to-end distance $R(N)$ and differes from it by a factor of order one. Similarly, one can also use the gyration radius of a chain fragment of length $s$, $S(s)$. Also, the average distance between two units separated by a contour length $s$, $R(s)$, can be applied to ring chains in a way similar to what have been done above for linear chains, provided that $s \ll N$.

\subsubsection{Equilibrium melts of rings}

A polymer in an array of obstacles is an illustrative but a bit artificial example. However, it is easy to see that the phenomenon of topological polymer phases should be much more wide-spread. As we have seen, ring chains can only have conformations that conserve the initial topological state of the ring, which makes them very different from linear chains in certain situations. Extended conformations of polymers (i.e. ideal and swollen coils) are typically unknotted and large-scale statistics for unknotted rings are asymptotically the same as for linear chains up to a numerical factor of order 1. (Actually, this is also true for rings with non-trivial knots, provided that the knot topology stays fixed as $N \to \infty$.) On the other hand, condensed conformations of polymer chains (equilibrium globules and melts) are very entangled, and, as a result, conformational statistics of rings and linear chains in these systems are radically different. 

As we already know from section 1.4, the structure of melts and globules are similar, although the melts are somewhat simpler systems since one does not need to bother about boundary effects. Therefore, for more than 30 years since the pioneering paper \cite{cates} the archetypical system to study topological states of polymer chains is the melt of {\it nonconcatenated polymer rings}. Consider a set of closed polymer chains (rings) which are prepared in such a way that each ring forms a trivial knot and different rings are not concatenated. Then prepare a melt of such chains (by evaporating the solvent, for example) and study equilibrium conformations of the chains. Clearly, Flory theorem will not work for this system, as there will be {\it topological repulsion} between the chains. 

To understand that, imagine for a moment that chains are phantom (can cross each other). Then the Flory theorem will apply, similarly to the linear chain case, and the typical conformations of the rings will be ideal. According to \eq{density} the average concentration $n$ of monomers of a ring inside the volume spanned by it, is very small, proportional to $N^{-1/2}$. Therefore, this same volume will be penetrated by many other chains, the corresponding {\it overlap parameter} scaling as $c^{-1} \sim N^{1/2}$. Clearly, if an equilibrium phantom ring overlaps with of order $N^{1/2}$ other rings, it will be concatenated with at least some of them in a typical conformation.

Therefore, if we now return to the behavior of phantom chains which are forced to preserve the fact they are non concatenated, their conformations must get more contracted so that the overlap parameter is reduced. Will this contraction be enough to make the rings compact, so that their gyration radii scale as $N^{1/3}$? In the original paper\cite{cates} by Cates and Deutsch authors constructed an estimate for the free energy of a ring as a function of its spatial span in a way similar to the classical Flory theory of swollen linear chains (see section 1.3). According to this theory the equilibrium size of the ring is determined by a compromise between entropical cost of contraction (indeed, the most entropically favorable conformations are ideal ones) and the free energy cost associated with topological repulsion of the rings, which tries to make them more compact. The entropy loss is estimated in \cite{cates} to be equal to the entropy loss of an ideal chain confined in a pore of size $R$, $F_{entr} \sim N T a^2/R^2$. The cost associated with topological repulsion is estimated to be proportional to the number of neighboring rings penetrating into the same volume as the ring under investigation, giving rise to a term of order $F_{topol} \sim T R^3/N a^3$ (topological interaction is due to reduction of the phase space because of non-phantomness; therefore it is intrinsically of entropic nature, thus the proportionality of the free energy to temperature $T$). Combined, these terms constitute the free energy of the following form
\be
\frac{F}{T} \sim \frac{Na^2}{R^2} + \frac{R^3}{Na^3}.
\label{f_cates}
\ee
Minimization of this free energy with respect to $R$ gives:
\be
R \sim N^{2/5} a,
\ee
i.e. rings get contracted compared to linear chains (recall that $R \sim a N^{1/2}$ for linear chains in the melt), but equilibrium conformations are still more extended than in a compact liquid-like globule with $R \sim N^{1/3}$.
Early computational works were in a good agreement with the critical exponent 2/5 \cite{muller86}, but in later years the bulk of simulation data, including both Monte-Carlo lattice dynamics\cite{vettorel09} and off-lattice molecular dynamics\cite{halverson11,halverson11_2} convincingly demonstrated that in the limit $N\to \infty$ the true critical exponent is actually 1/3, i.e. the rings do become compact in the end of the day. However, the crossover to this behavior is very slow, and there is a wide range of intermediate chain lengths for which the contraction can be approximated by the scaling exponent 2/5. 

In several theoretical papers\cite{sakaue,sakaue2,obukhov,grosberg13,ge16} attempts have been made to modify the original approach\cite{cates} and to reproduce the existing numerical data. All these attempts share some common features. First, there is still no theory allowing to obtain a Hamiltonian for topological interactions from the first principles, so all the authors rely on speculations of Flory-like (``write down a reasonably convincing free energy of the chain as a function of its spatial size and minimize it'') or renormalization-group-like nature. Second, the basic physical insight, namely, that the ring conformations are controlled by the interplay of topological interactions which try to make conformations more compact, and single-chain entropic effect which tend to make them more extended, is sound and remains the same in all approaches. Third, the resulting chain conformations are asymptotically compact but corrections to asymptotic behavior $S \sim N^{1/3}$ decay very slowly, as a power law of $N$ with some small negative power. Importantly, if $N$ is rescaled in the units of the so-called {\it entanglement length} $N_e$, then the $S(N)$ curves for chains with different microscopic structure seem to collapse onto a single master curve over three orders of magnitude from $N/N_e \approx 0.1$ to $N/N_e \approx 150$\cite{halverson12}. The notion of entanglement length $N_e$ was first introduced in the {\it reptation theory} for the dynamics of melts of linear chains\cite{doi_edwards, degennes,gr_khokhlov,rubinstein} as a measure of chain length at which entanglement effects become important. Entanglement length is typically about 50-500 monomer units; however, the question of whether {\it topological} entanglement length (relevant for the problem of equilibrium ring conformations) and {\it dynamical} entanglement length (relevant for the problem of reptational dynamics) are the same is a matter of some controversy in the literature.

Finally, both theoretical works \cite{obukhov,grosberg13,ge16} and simulations \cite{rosa_everaers14,mirny_nechaev} seem to show that ring conformations are self-similar in a sense that a part of the long ring behaves essentially like a shorter ring: conformational statistics (for example, distribution of $S(s)$) of a fragment of given length $s$ belonging to a longer ring of length $N$ is almost independent of $N$ up to $s \approx N/2$ or even further. As a result, compact conformations of very long rings can be seen as fractal objects with fractal dimension $d_f = (1/3)^{-1} = 3$. Shorter rings form approximately fractal structures with $d_f$ slowly converging to 3 with the growing length of the ring (with $d_f \approx 5/2$ being a reasonable approximation at intermediate sizes). Note that this structure is radically different from conformation statistics of equilibrium globules of linear polymers: for linear polymers the total size of the structure also behaves as $S(N) \sim N^{1/3}$ but conformations are not fractal and spatial size of chain fragments is described by a more complicated formula \eq{radius_globule}. 

\begin{figure}[!t] \centering
\includegraphics[width=\textwidth]{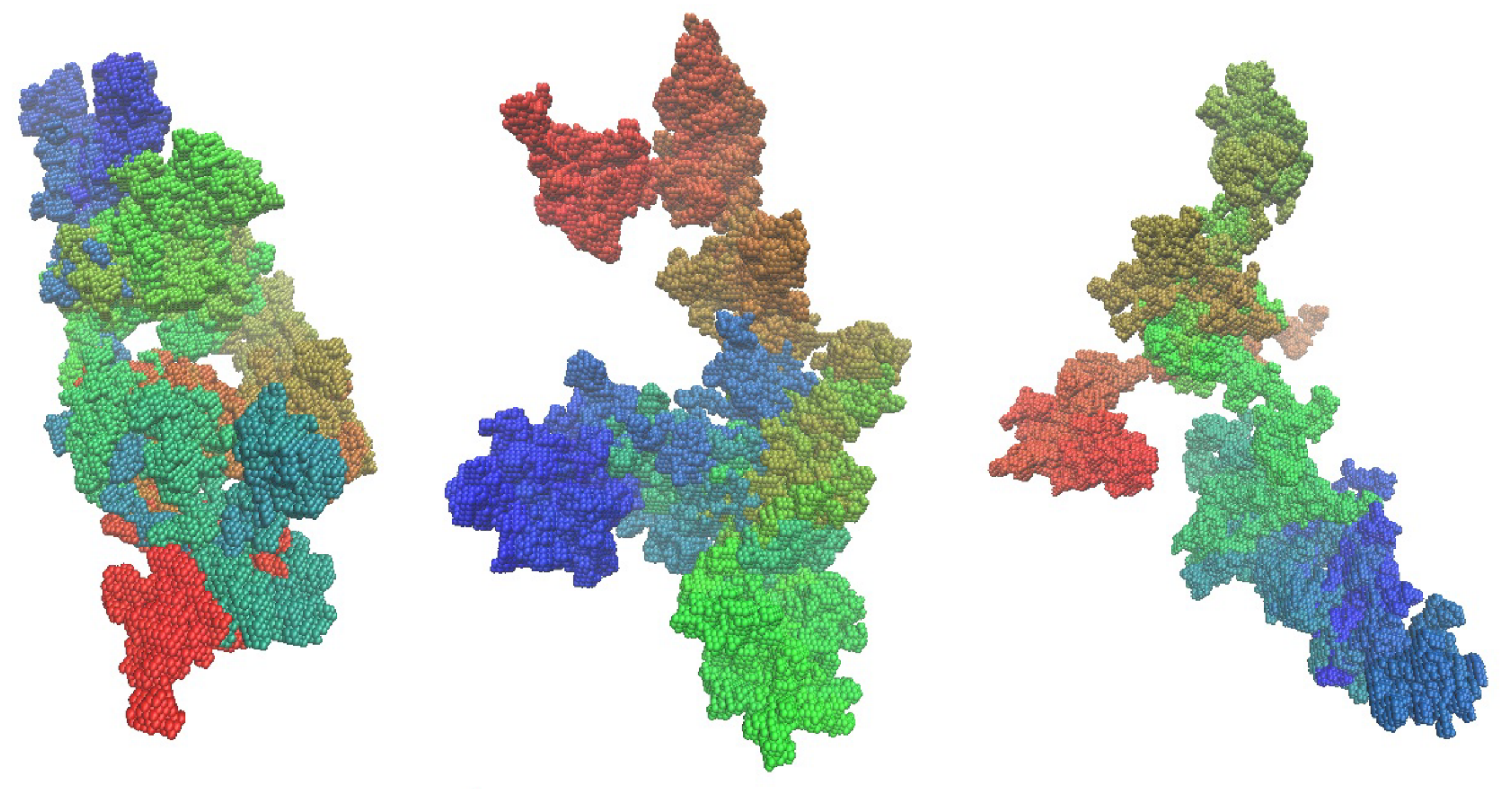}
\caption{Examples of typical shapes of compactized topological states of long polymer chains. The picture presents conformations of linear polymer chains generated by the algorithm of conformation-dependent polymerization\cite{tamm15,nazarov_phd}, the large-scale statistical properties of conformations obtained with this method are indistinguishable from those of compactized rings in a melt.
Chains are colored from red on the one end to blue on the other, so that the domain structure is clearly seen. The figure is courtesy of L.I.~Nazarov.}
\label{fig:3}
\end{figure}

Let us now go beyond the fractal dimensionality of the packing and discuss two other interesting characteristics of the ring conformations: fractal dimensionality of the surface and return probability. Indeed, for extended polymer conformations with $d_f < 3$ almost all monomer units are in contact with monomer units belonging to other polymer chains. It is not necessarily true in the case of compact conformations for which $d_f = d = 3$. Consider one very long ring in such a conformation. One may expect that some of its monomers are deep inside the bulk of the compact structure and are not exposed to other chains, but some are not. If the number of these surface monomers grows as a power law
\be
N_{surf} \sim N^{d_s/d},
\label{surface}
\ee 
then we will call $d_s$ a {\it surface exponent} of the structure. For example, a liquid droplet, an equilibrium globule, or any other structure with finite surface tension will have a surface exponent $d_s = d-1 = 2$. However, there is no surface tension between two compactized rings in a melt (indeed, they are formed by chemically similar monomer units, and therefore intra-chain and inter-chain volume interactions are exactly the same) and thus there is no a priori reason to expect $d_s = 2$. Indeed, numerical simulations show that compactized ring states typically have very developed surfaces (see \fig{fig:3}). On the other hand, it seems to be well-established that it is strictly less than 3 and results obtained by various authors\cite{halverson13} show exponents ranging from $d_s = 2.5$ to $d_s = 2.9$. There exist a theoretical derivation of $d_s \approx 2.75$\cite{grosberg13} but it is not free of somewhat arbitrary uncontrolled approximations. 

The second structural characteristic, which turns out to be closely related to the ring exponent, is the {\it return probability} defined as a probability for two monomer units separated by a contour distance $s$ to be situated close to each other in space (``close'' here can be defined as ``at a distance smaller than the distance between two adjacent monomers along the chain $a$'', although the corresponding critical exponent is independent of exact definition of closeness). Now, if this return probability $P(s)$ decays as a power law for large $s$, then one can define a critical exponent $\alpha$:
\be
P(s) \sim s^{-\alpha}
\label{prob}
\ee
Na\"{i}vely, one can expect $\alpha = d/d_f$. Indeed, assume one monomer unit is fixed at the origin. Then the volume accessible for a monomer unit at a contour distance $s$ from the fixed one is proportional to $R(s)^{d} \sim s^{d/d_f}$. If the probability of getting into each of these points (including the origin) is roughly the same, one gets
\be
P(s) \sim \frac{a^d}{(R(s))^d} \sim s^{-d/d_f}.
\ee
This result, however, is true only for ideal polymer conformations, where the return exponent indeed equals $\alpha = 3/2$ in 3d. Indeed, for interacting chains the probability of the distance to be close to zero is significantly different from probability of it being equal to any other accessible value, which breaks the above reasoning. For swollen linear chains one can show that return probability equals
\be
\alpha = d \nu + \gamma,\;\; \alpha \approx 2.1 \text{ in 3d},
\ee
where critical exponents $\nu$ and $\gamma$ are defined in \eq{critical}. For compact ring conformations the return exponent turns out to be closely related to the fractal dimensionality of surface $d_s$. Indeed, consider a chain fragment of length $s$. According to \label{surface} (and to the general principle that compact ring conformations are self-similar fractal structures) it forms a compact domain with a surface area $s^{d_s/d}$. If we want a monomer in this fragment to be in contact with a monomer outside the fragment (i.e. at a contour distance more than $s$), it should belong to the surface of this compact domain. Therefore, the total number $n(s)$ of monomers inside the domain in contact with monomers outside the domain can be estimated as
\be
n(s) \sim z s^{d_s/d} \sim s \int_s^{\infty} P(x)dx \sim s^{2-\alpha},
\label{db2}
\ee
where $z$, a typical number of contacts per surface monomer, is a number of order unity. Therefore, the following scaling relation holds:
\be
\alpha = 2 - \frac{d_s}{d}.
\label{db}
\ee
As a result, typical values of $\alpha$ for compact ring conformations lie between 1 and 1.2, which is radically different from the value of 1.5 which is observed in melts of linear chains and in equilibrium globules (for $s < N^{2/3}$). This untypically low value of return probability exponent has been well documented in computer simulations of polymer rings\cite{halverson13}. Even more importantly, similar values of return probability have been observed experimentally for the statistics of chromosome packing in many eukaryotes\cite{lieberman}, which gave rise to the hypothesis that chromosome are packed into domains reminiscent of compactized ring conformations (see section 1.6.5 for more details). 

\subsubsection{Ring melts and annealed randomly branched structures} 

As mentioned above, there is still no well-established first-principle theory describing conformations of polymer rings in a melt. However, there is growing computational evidence connecting the properties of the melts of rings to those of annealed randomly branched structures. That is to say that not only a ring in an array of immovable obstacles\cite{khokhlov85} discussed in section 1.6.1, but also a ring surrounded by other similar rings form a sort of approximately double-folded structure with a randomly branched skeleton. This conclusion is not a priori obvious (indeed, one can easily imagine one tree forming a wide open loop, and another one protruding through this open space), but experimental evidence seems to support it more and more. For example, Smrek and Grosberg have shown\cite{smrek} that if one spans a minimal surface onto a ring in a melt, its area grows almost exactly linearly with the length of the ring $N$ for $N \gg N_e$, indicating that conformations are indeed double-folded on the large scale. In turn, Rosa and Everaers\cite{rosa_everaers14} provided an extensive numerical comparison of conformational statistics for nonconcatenated rings and annealed randomly branched polymers in a melt and found, after proper rescaling, almost exact similarity between the properties of the two systems.

The most promising recent advances in the theory of ring melts\cite{obukhov,grosberg13,ge16} also rely on the analogy between ring conformations and those of annealed randomly branched trees. However, let us emphasize that while the behavior of  ideal randomly branched chains is comparatively easy to study, and the critical exponent $\nu = 1/4$ for this system is known for a very long time\cite{zs49}, the behavior of {\it interacting} annealed trees is in itself a very murky subject with very little exact results known. Recently, there have appeared several papers with comprehensive systematization of numerical information\cite{rosa16,rosa17} and theoretical predictions\cite{everaers17}concerning the behavior of randomly branched trees in different conditions, and we address interested readers to these papers for more details.

\subsubsection{ Fast collapse of a linear polymer chain: crumpled globule}

As described in previous two subsections, topological states seem to be a peculiarity of systems of ring polymer chains as opposed to linear chains. However, such a statement is more than a bit bizarre in the afterthought. The difference in conformation statistics of linear and ring chains is apparent on the scale of $N_e$ monomer units which remains finite as the total length of a chain $N$ goes to infinity. How a relatively short fragment of an infinitely long chain could possibly know whether it belongs to a linear or circular chain? 

To clarify this subject consider a following thought experiment. Let us start with an equilibrium melt of nonconcatenated rings and suddenly at time $t=0$ cut all the rings open (that is to say, remove one bond per ring without changing coordinates of any monomer). What will happen? As time goes on, the system, stripped of its topological constraints, will eventually relax to the state of an equilibrium melt of linear chains, with chains adopting ideal Gaussian conformations. However, this relaxation will take quite a long time of order $N^3$ (see section 3). Meanwhile, at times much shorter than this relaxation time all properties of the melt of linear chains will largely resemble those of the melt of rings. That is to say, topological conformations, which in equilibrium can only exist in systems of ring chains, may still be observed in systems of linear polymers as {\it long-living metastable states}, with lifetime of these states diverging as the length of the chains goes to infinity. 

\begin{figure}[!ht] \centering
\includegraphics[width=0.7\textwidth]{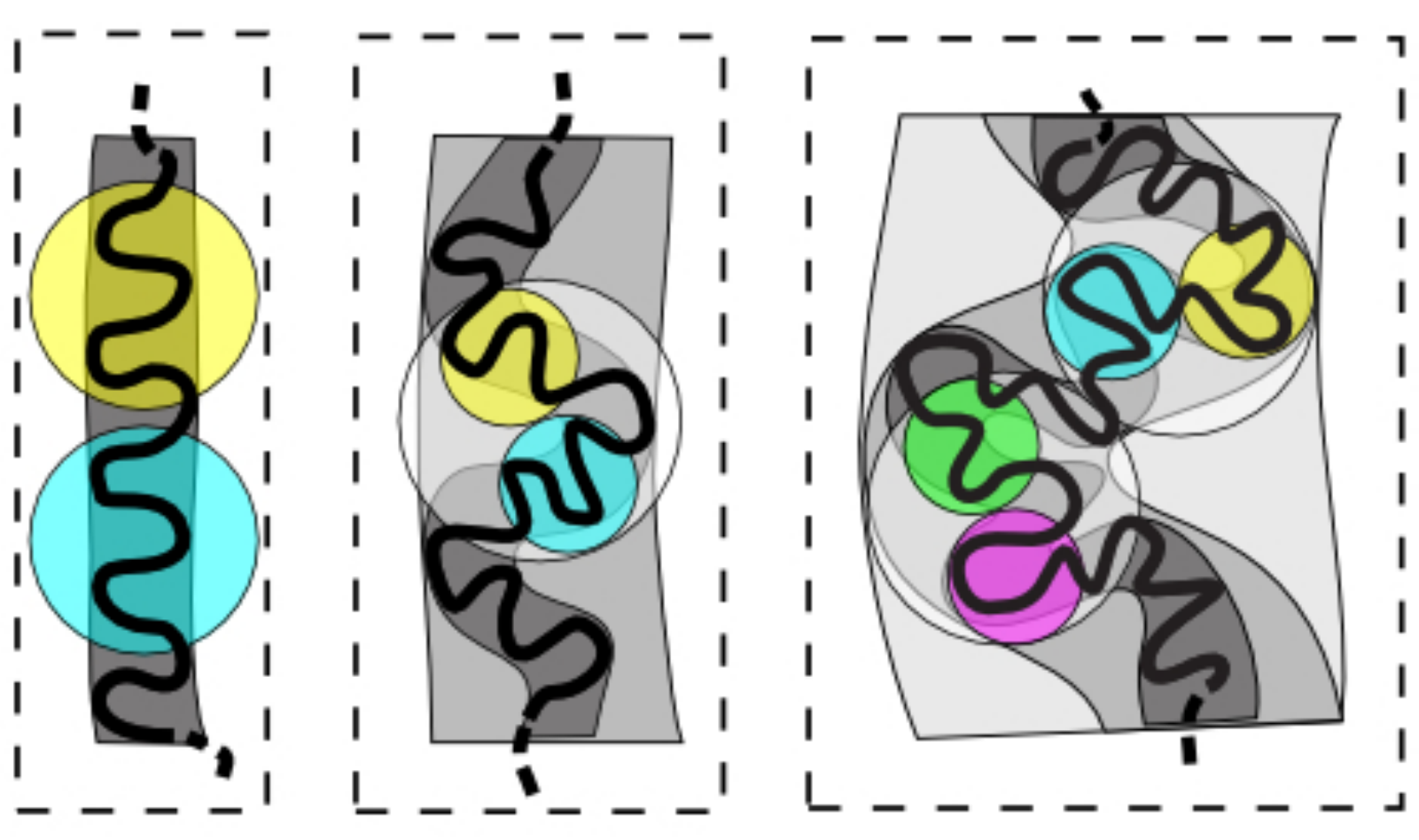}
\caption{The idea behind the concept of a crumpled globule. Rapidly collapsing polymer chain forming an hierarchy of folds (crumples) on larger and larger scales. Figure adopted from \cite{nazarov15}.}
\label{fig:4}
\end{figure}

First example of such a metastable state was suggested almost 30 years ago in a seminal paper\cite{gns88} where the
authors considered intermediate stages of the collapse of a linear chain from a swollen coil to a condensed globule under a sudden and drastic change of solvent quality. They argued that, since typical conformations of swollen coils are unknotted\footnote{More precisely, if one connects the ends of a swollen linear chain by a straight line and considers topological properties of a resulting circular structure, it will usually be a trivial knot. Other simple knots may also occur, but it is important that a typical knot will not become more complicated as the length of the chain diverges.}, this should remain true on the early stages of collapse as well, provided that the collapse is fast enough. The resulting unknotted condensed structure, which the authors of\cite{gns88} called a {\it crumpled globule}, can be though of as an hierarchy of small crumples combining into larger and larger crumples (see \fig{fig:4}). Such a structure is obviously very reminiscent of the condensed ring conformation described above: it is fractal with fractal dimensionality 3, it has, contrary to an equilibrium globule, a well defined domain structure (subchains of shorter length also form compact sub-globules with fractal dimensionality 3) and, most importantly, the physical reason for the formation of this structure is of the similarly topological nature as in the case of the ring melt described above: the need to preserve a trivial knot topology on the scale of the whole chain. 

It was shown in recent years\cite{schram,kos} that the original idea\cite{gns88} does not work exactly as invented: even in a very poor solvent the collapse is not fast enough to give rise to a proper crumpled globule. In fact, collapse goes on as a consecutive coalescence of larger and larger droplet-like sub-globules\cite{bunin_kardar}, with later stages happening much slower than earlier ones, so that small sub-globules have time to partially relax before bigger ones are even formed. As a result, on all intermediate stages the non-equilibrium globule exhibits properties somewhere in between equilibrium globule and purely crumpled globule. 

However, formation of a crumpled globule almost exactly as predicted in\cite{gns88} turned out to be possible for a case when a coil is collapsing in a strong external field \cite{lieberman,mirny11}. The resulting simulations of crumpled globules turned out to be highly instructive for the interpretation of the new experimental data on the spatial packing of chromosomes. 

\subsubsection{Spatial packing of chromosomes} 

If one wants to prepare a state mimicking the topological state of nonconcatenated ring melt in a system of linear chains, there is another simple possibility. Instead of rapid condensation from a diluted solution of coils described in the previous subsection, one can start with a condensed phase consisting of chains which are somehow prepared to be in a state with no entanglements or knots. Seemingly, this is exactly what happens to chromosomes during the cell cycle in eukaryotic cells. Indeed, during mitosis (cell division) chromosomes are separated from each other, and each chromosome forms a cylindrical globule with a very regular and unentangled internal structure (which looks like a sequence of chamomile-like structures glued to each other). When the division is finished, chromosomes are released into a newly formed cell nucleus, which, during the interphase (the stage of the cell cycle between divisions) amounts to a concentrated solution of chromosome chains. 

That is why the crumpled globule was suggested as a possible model for the structure of chromosomes as early as in 1993\cite{grosberg93}. There is a long history of evidence supporting the idea that crumpled globule and nonconcatenated ring conformations might be relevant to the chromosome structure and that chromatin (the complex DNA-protein fiber that constitutes chromosomes) is packed inside the nucleus in a way rather different from linear polymer melts. The striking difference between melts obeying Flory theorem and interphase chromatin consists in the fact that in the latter case each chromosome takes a distinct separated spatial territory\cite{territories}, similarly to what one might expect in a melt of rings. 
Also, it has long been argued \cite{sikorav94} that potential knots and entanglements would make the functioning of 
all relevant biological machinery much more complicated and dependent on auxiliary enzymes such as topoisomerase.

The idea that chromatin folding can be modelled as a melt of linear chains which are prepared in a highly ordered state and remain in a metastable non-equilibrium state at biologically relevant times was first suggested and checked in computer simulations in\cite{rosa08} where authors not only observed the formation of chromosome territories, but quantitatively reproduced the $R(s)$ dependence for chromosomes obtained in FISH experiments\cite{fish}. Further evidence in favor of topological explanation of eukaryotic chromatin structure comes from the fact that for typical mammal cells the relaxation time of the chromatin melt is estimated by reptation theory to be of order $500$ years, which is obviously much larger than the typical time scale of cell division\cite{rosa08}. Additional factors may also slow down the relaxation, including the presence of telomers at the ends of the chromosomes which might suppress reptation\cite{telomers} and the fact that some parts of chromatin are attached to the inner surface of the nucleus (lamina).

Recently, crumpled globule and other topological models of chromosome packing got a significant boost after the development of the Hi-C method\cite{dekker}, which allows to measure the ensemble averaged contact probability between any two chromosome loci $P_{ij}$\cite{lieberman}. The resulting $N\times N$ matrices of probabilities, so-called Hi-C maps, typically have a very rich fine structure as functions of both $i$ and $j$, but average contact probability at a given contour distance
\be
P(s) = \langle P_{i,i+s}\rangle_i
\ee 
indicates a clear power law decay as a function of $s$ with exponent between 1.0 and 1.2 for the interphase chromosomes of most eukatyotic organisms \cite{lieberman,zhang12}. This is in good agreement with expected behavior of compactized rings \eq{prob},\eq{db}\footnote{The only exception is yeast\cite{yeast} cells that have short chromosomes which manage to relax to equilibrium at biological time-scales\cite{rosa10}. In that case a decay with exponent 1.5 typical for an equilibrium globule is observed \eq{radius_globule}}. 

Let us emphasize that biology, as always, is much richer in fine details than a simple synthetic polymer system where chains of identical monomer units are floating in a macroscopically large volume of homogeneous solvent. Theory of the spatial organization of chromosomes is a very rapidly developing field. It is clear that factors like non-homogeneity of chromatin fiber, which can be in different internal states\cite{nazarov15, jost}, formation of site-specific reversible bridges and bonds\cite{langowski1,langowski2,nicodemi,ulyanov}, sliding of such bonds via the so-called loop extrusion mechanism \cite{fudenberg16}, active forces, etc. seem to play very important role in formation and stabilization of the fine and complicated chromosome structure. However, it is now widely accepted that chromatin is in a substantially non-equilibrium state, and because of that topological effects are crucial for understanding of its structural properties.

\section {Unentangled polymer dynamics} 

\subsection{Rouse model}

We now have enough information about equilibrium properties of polymer systems to address the main topic of this chapter, namely, the dynamics of polymer systems. As in the discussion of the equilibrium properties, we will concentrate here on the properties of single chains as opposed to, for example, stress-strain curves of polymer materials as a whole. 

Before constructing any models, let us consider the most important factors which influence the dynamics of a polymer in a solvent. First of all, there are properties of the medium in which polymer moves. We will start with the simplest case, that of a polymer in a viscous Newtonian fluid, which exhibits linear viscous response to external force and is in thermal equilibrium so that the fluctuation-dissipation theorem is respected. However, it often makes sense, especially in biological applications, to consider polymers positioned in a liquid with more complicated viscoelastic properties \cite{theriot10a}. Indeed, due to spatial inhomogeneity, molecular crowding and other factors biological liquids such as cytoplasm and nucleoplasm are known to exhibit mechanical properties different from those of a simple Newtonian liquid. An important example of non-Newtonian liquid in thermal equilibrium is a liquid in which a probe particle undergoes subdiffusion subject to the so-called generalized Langevin equation \cite{metzler_review}. We will touch this subject briefly later in this section. Even richer behavior can probably be observed in dynamics and steady states of polymer chains surrounded by a non-equilibrium media such as granular matter or active matter, but very little is known on this subject yet. 

On the polymer physics side, there are several factors which are familiar to the reader from our discussion of the equilibrium properties. First, there is chain connectivity itself: the fact that monomer units (or beads) adjacent along the chain cannot go too far from each other. Second, there are volume interactions: one expects that a swollen chain will move differently than an ideal one. Third, the non-phantomness of the chain is crucial for understanding of the dynamics of melts. On top of these three factors there are hydrodynamic interactions: a moving monomer unit disturbs the solvent around it and the resulting movement of the solvent affects other monomer units. These interactions decay very slowly with distance, and, as we will see in section 2.4, they can drastically influence the properties of the chain. 

However, as a start it is natural to consider the simplest possible model, which preserves, out of the four factors mentioned above, just the chain connectivity. This is exactly the approach taken by Rouse\cite{rouse} which constitutes the celebrated ``Rouse model'' of polymer dynamics. The idea of the model is to take an ideal polymer chain and allow independent thermal forces to act on its monomer units.

\subsubsection{Formal definition of the Rouse model}

In terms of the standard beads-and-springs model the formal definition of the Rouse model is as follows. As usual, the chain conformation is defined by a set of coordinates $\ve r_n(t), n = 0...N$ which are now explicitly time-dependent. The Langevin equation of motion for a bead reads:
\be
m \frac{d^2 \textbf r_n}{dt^2} = -\xi \frac{d \textbf r_n}{dt} - \frac{\partial}{\partial \ve r_n} U(\textbf r_0, ..., \textbf r_N) + \textbf F_n,
\label{langevin_1}
\ee
where $m$ is the mass of the bead, and the first term on the right hand side is a friction force, $\xi$ being the friction coefficient between the bead and the solvent; the second term, where $U(\ve r_0, ... \ve r_N)$ is given by \eq{energy_ideal}, is a force acting on a bead from all other beads of the chain, and $\ve F_n$ is a random thermal force. In what follows we will restrict ourselves to the overdamped regime, when the left hand side of \eq{langevin_1} is negligibly small and the equation is reduced to a first order differential equation in time. Then, substituting \eq{energy_ideal} into \eq{langevin_1} one gets the following set of 
$N+1$ linear differential equations:
\be
\left\{
\begin{array}{l}
\xi \frac{d \textbf r_n}{dt} = k(\textbf r_{n+1} + \textbf r_{n-1} - 2\textbf r_n) + \textbf F_n, \qquad n = 1, ..., N-1 \medskip \\
\xi \frac{d \textbf r_0}{dt} = k(\textbf r_{1} - \textbf r_0) + \textbf F_0 \medskip \\
\xi \frac{d \textbf r_N}{dt} = k(\textbf r_{N-1} - \textbf r_N) + \textbf F_N
\end{array} \right.,
\label{disrouse}
\ee
where we used a short-hand notation $k = d T/a^2$ for the spring stiffness. 

As for the random force, $\textbf F_n (t)$, it is natural to assume it to be normally distributed with 
\be
\langle \textbf F_n(t) \rangle = 0, \; \; \langle \textbf F_n(t) \textbf F_n(t') \rangle = 2d T \xi \delta(t - t'),
\label{meanf}
\ee
where the zero mean corresponds to the assumption that the solvent is at rest (no macroscopic flow and no hydrodynamic interactions) and the form of the force correlator is dictated by fluctuation-dissipation theorem for the case of instantaneous friction force as written in the first term on the right-hand side of \eq{langevin_1}. Moreover, a natural first approximation is to assume that random thermal forces acting on different beads are independent, and therefore
\be
\langle \textbf F_n(t) \textbf F_m(t') \rangle = 2d T \xi \delta(t - t') \delta_{n,m}.
\label{cov}
\ee
Physically, this independence assumption means that the typical bead-to-bead distance $a$ is much larger than the correlation length of the solvent itself. Note that this is always possible to achieve provided the chain is long enough: as we mentioned in section 1.1, $a$ is a discretization parameter and the choice of $a$ is in fact arbitrary provided that it is larger than the persistence length $l$ of the underlying chain. 

Finally, it is instructive to define the Rouse model for the case of a continuous chain with uniform flexibility, as introduced in section 1.1. To obtain the continuous version of \eq{disrouse} one should notice that the expression on the right-hand side of the \eq{disrouse} for $n=1,...,N-1$ is nothing but the discretized version of second derivative by $n$. Therefore, the continuous analogue of the Langevin equation for the string dynamics should be written on $\ve r(s, t)$, where $s \in (0, L)$ as follows:
\be
\xi \frac{\partial \textbf r (s, t)}{\partial t}= k\frac{\partial^2 \textbf r(s, t)}{\partial s^2} + \textbf F(s, t), \qquad s \in (0, L)
\label{conrouse}
\ee
with force correlators
\be
\langle \textbf F(s,t) \textbf F(s',t') \rangle = 2d T \xi \delta(t - t') \delta(s-s'),
\ee
where the first term in the right-hand side is nothing but the derivative of the potential \eq{persistent}. For simplicity of presentation we use the same notations $\xi, k, \ve F$ for parameters (friction coefficient, chain stiffness, and thermal force, respectively) of both the discrete \eq{disrouse} and continuous \eq{conrouse} versions of the Rouse model. In what follows it is always clear which particular version is considered in any given moment, so it should not, in our opinion, result into any confusion. Boundary conditions at $s=0,L$ are easy to obtain from the last two equations of \eq{disrouse} and read
\be
\frac{\partial \textbf r(s, t)}{\partial s} \Bigg|_{s=0} = \frac{\partial \textbf r(s, t)}{\partial s} \Bigg|_{s=L} = 0.
\ee 

The equation \eq{conrouse} is, of course, nothing but an equation for thermal fluctuations of a uniform string, and as such is identical to the celebrated Edwards-Wilkinson equation\cite{krapivsky_book}. The only difference is that we consider a vector function $\ve r(s,t)$ and not a scalar one, as it is usually done in the theory of fluctuating surfaces.

\subsubsection{Solution of the Rouse equation. Relaxation modes}

Both the discrete \eq{disrouse} and the continuous \eq{conrouse} versions of the Rouse equation are exactly solvable by Fourier transform, or, in theoretical mechanics terminology, by changing variables to the so-called {\it normal modes}. Let us start with the discrete case. In order to diagonalize \eq{disrouse},
we perform the Fourier transform $\{\textbf r_n\}_0^N \to \{\textbf u_p\}_0^{N}$ as follows:
\be
\textbf r_n(t) = \sum_{p=0}^{N} \textbf u_p(t) \alpha_p^{(n)}; \quad \textbf u_p(t) = \sum_{n=0}^{N} \textbf r_n(t) \alpha_p^{(n)},
\label{Fourier}
\ee
where coefficients $\alpha_p^{(n)}$ are chosen to make the formulae for direct and inverse Fourier transform as simple as possible, and read
\be
\alpha_p^{(n)} = 
\sqrt \frac{2-\delta_{p,0}}{N+1} \cos \frac{p\pi(n-1/2)}{N}, \; p=0,...,N.
\label{alphas}
\ee

In these normal coordinates $\ve u_p(t)$ the spring potential adopts the following diagonal form:
\be
U(\ve u_0,\textbf u_1, \textbf u_2, ..., \textbf u_{N}) = \frac{1}{2} \sum_{p=1}^{N} \kappa_p \textbf u_p^2
\label{normal_potential}
\ee
where
\be
\kappa_p = 4k\sin^2 \frac{p\pi}{2N}
\label{disk}
\ee
As a result, the Langevin equations \eq{disrouse} decouple, resulting in the following set of equations:
\be
\begin{array}{rll}
\displaystyle \xi \frac{d \textbf u_0}{dt} &= &\textbf {f}_0, \medskip \\
\displaystyle \xi \frac{d \textbf u_p}{dt} &= & -\kappa_p \textbf u_p + \textbf {f}_p, \text{ for } p=1,...,N 
\end{array}
\label{OU}
\ee
where $\textbf {f}_p$ is the Fourier image of the random force
\be
\textbf {f}_p = \sum_{n=0}^N \textbf F_n \alpha_p^{(n)},
\ee
which, being a linear combination of normally distributed independent random forces, is itself normally distributed with correlators
\be
\qquad \langle\textbf f_p(t) \textbf f_{p'}(t)\rangle = 2d T\xi \delta_{p,p'}\delta(t-t').
\ee
The zero-th mode $\textbf u_0(t)$, which describes the position of the center of mass of the chain,
\be
\textbf u_0(t) = \sqrt\frac{1}{N+1}\sum_{n=0}^{N} \textbf r_n(t) = \sqrt {N+1} \textbf r_{c.m.}(t)
\ee
is therefore subject to standard Brownian motion with explicit solution
\be
\begin{array}{l}
\displaystyle \textbf u_0(t) = \textbf u_0(0) + \xi^{-1}\int_0^t \textbf f_0(\tau) d\tau; \qquad p = 0, \medskip \\
\displaystyle  \langle \left(\textbf u_0(t)-\textbf u_0(0)\right)^2 \rangle =2\frac{dT}{\xi} t
\end{array}
\label{zeroth}
\ee
(here and below, if not otherwise stated, angular brackets $\langle...\rangle$ designate the average over different realizations of the random force). It is clear that at very long time scales the chain as a whole moves similarly to its center of mass. Indeed, the distance travelled by the center of mass is, according to \eq{zeroth}, unbound while chain connectivity prevents beads from moving to far away from the center of mass. Therefore, at very long times the chain as a whole undergoes simple diffusion with a diffusion coefficient inversely proportional to the total number of beads, $N+1$:
\be
\langle \left(\textbf r_{c.m.}(t)-\textbf r_{c.m.}(0)\right)^2 \rangle = \frac{\langle \left(\textbf u_0(t)-\textbf u_0(0)\right)^2 \rangle}{N+1} = \frac{2dT}{\xi (N+1)} t =2Dt.
\label{discm}
\ee
The physical meaning of this result is quite clear: since the friction forces acting on the beads are independent, the effective friction coefficient for the whole polymer coil is $(N+1)$ times the friction coefficient for a single bead $\xi$.

Now, the dynamics of $\ve u_0$ gives a dominant contribution into the coil movement only in the long time limit. To understand coil movement on shorter timescales, as well as to estimate the actual timescale at which simple diffusion becomes dominant, one needs to analyze the dynamics of all other normal coordinates. Their dynamics is described by a set of independent Ornstein-Uhlenbeck equations \eq{OU} whose formal solution reads
\be
\textbf u_p(t) - \textbf u_p(0) = \frac{1}{\xi}\int_0^t \textbf f_p(\tau) \exp\left(-\frac{t-\tau}{\tau_p}\right) d\tau; \qquad p \ge 1.\notag\\
\label{solt}
\ee
Since the right-hand side of \eq{solt} is a linear function of the white noise $\textbf {f}_p$, it is itself normally distributed with zero mean and dispersion which is easy to calculate directly:
\be
\langle \left(\textbf u_p(t)-\textbf u_p(0)\right)^2 \rangle = \frac{d T}{\kappa_p}\left(1 - \exp\left(-\frac{2t}{\tau_p}\right)\right); \qquad p \ge 1
\label{dispu}
\ee
where $\tau_p$ is the relaxation time of the $p$-th Rouse mode,
\be
\tau_p = \frac {\xi} {\kappa_p} = \frac{\xi a^2}{4d T}\sin^{-2} \frac{p\pi}{2N}.
\label{distimes}
\ee
The shortest relaxation time
\be 
\tau_N = \frac{\xi a^2}{4d T}=\tau_0
\label{microtime}
\ee
has a meaning of the typical time needed for a single free bead to diffuse a distance of order $a$ and works as a natural microscopic timescale. In what follows we prefer to denote it $\tau_0$ rather than $\tau_N$ to emphasize that it is a microscopic variable independent of the chain length $N$. In turn, transition to simple diffusion regime is controlled by the largest relaxation time $\tau_{\max}(N)$
\be 
\tau_{\max}(N) = \tau_1 = \tau_R = N^2 \frac{\xi a^2}{\pi^2 dT} =\frac{4}{\pi^2} N^2 \tau_0,
\label{rouse_time}
\ee
we use $\tau_{\max}(N)$ to designate maximal relaxation time of the chain in all model, while $\tau_R$ is its particular value for the Rouse model, known as Rouse time.

For any given $p \ge 1$, at times larger than $\tau_p$, the mean-square displacement of
the normal coordinates converges to its value predicted by equipartition theorem:
\be
\langle \left(\textbf u_p(t)-\textbf u_p(0)\right)^2 \rangle_{t \to \infty} = \frac{d T}{\kappa_p} = \frac{a^2}{4} \sin^{-2} \frac{p\pi}{2N}.
\label{normal_long-time}
\ee
On the other hand, for $t \ll \tau_p$ the mean-square displacement of $\textbf u_p$ grows linearly with time
as in normal diffusion, and the effective diffusion coefficient does not depend on $p$:
\be
\langle \left(\textbf u_0(t)-\textbf u_0(0)\right)^2 \rangle_{t \ll \tau_p} = \frac {2d T} {\xi} t.
\label{normal_short-time}
\ee
Therefore, for any given time $ t \in (\tau_0, \tau_1)$ all Rouse modes can be separated into two classes: modes with relatively small $p$ such that $t<\tau_p$ are not yet relaxed, their dynamics is independent of $p$ and described by \eq{normal_short-time}, while modes with larger $p$ are already thermalized and the dispersion of corresponding normal coordinate is described by thermal equilibrium \eq{normal_long-time}. 

Before providing any deeper analysis of the obtained solution, let us briefly outline the solution of the continuous version of Rouse equation\eq{conrouse}. In this case, instead of introducing a finite number of normal modes $\ve u_p(t)$, one introduces the infinite series of modes $\ve y_p(t)$ as follows:
\begin{align}
&\textbf y_p(t) = \frac{1}{\sqrt L} \int_{0}^{L} \ve r(s, t) \cos \frac{p\pi s}{L} ds; \qquad p = 0, 1, ... \notag\\
&\textbf r(s, t) = \frac{1}{\sqrt L}\textbf y_0(t) + \frac{2}{\sqrt N} \sum_{p = 1}^{\infty} \textbf y_p(t) \cos \frac{p\pi s}{L}
\label{contr}
\end{align}
For these modes, one obtains very similar results: the zero-th mode, which describes the movement of the center of mass of a chain, undergoes simple diffusion with diffusion coefficient inversely proportional to $L$. Other modes are once again subject to Ornstein-Uhlenbeck equations \eq{OU} with solutions \eq{solt}.The only difference is that continuous model does not have a well-defined microscopic cut-off, and the dispersion relation for the relaxation times is simpler:
\be
\kappa_p = p^2 \kappa_1; \qquad \tau_p = \frac{\xi}{\kappa_p}=\tau_R p^{-2}
\label{contimes}
\ee
(compare to \eq{distimes}). One can straightforwardly check that \eq{disk}, \eq{distimes}, and \eq{rouse_time} yield \eq{contimes} in the limit $p \ll N$. That is to say that the two approaches give exactly the same results for the dynamics of the chain at times much larger than the microscopic time of the discrete model, $t \gg \tau_0$. Thus the continuous Rouse model corresponds to simultaneously taking the limits $N \to \infty$ and $a \to 0$ (and, therefore, $\tau_0 \to 0$) in a way that leaves the Rouse time $\tau_R$ fixed.


\subsubsection{Single monomer dynamics in the Rouse model}

Consider now in more detail the dynamics of a single bead in the discrete version of the Rouse model. Note, first of all, that for any given time bead displacements are normally distributed: indeed, they are linear combinations of the normally distributed displacements of normal coordinates. As a result, the dynamics of each bead is fully described by its average
\be
\begin{array}{rll}
\langle (\ve r_n(t) - \ve r_n(0))\rangle & = &\sum_p \alpha_p^{(n)} \langle (\ve u_p(t) - \ve u_p(0))\rangle; \medskip \\
\overline{\langle (\ve r_n(t) - \ve r_n(0))\rangle} &= &0,
\end{array}
\ee
where the overbar here and below denotes averaging over initial conditions distributed according to thermal equilibrium, and the second equation follows directly from space isotropy; and mean-square displacement $x_n(t)$ defined as 
\be
x^2_n(t) = \overline{ \langle (\ve r_n(t) - \ve r_n(0))^2\rangle} =\sum_p (\alpha_p^{(n)})^2 \overline{\langle (\ve u_p(t) - \ve u_p(0))^2\rangle},
\label{displacement}
\ee 
where we took into account orthogonality of normal coordinates. Importantly, since $x_n(t)$ is averaged over equilibrium distribution of initial conditions, there is no dependence on the moment when we start observation, only on its duration $t$.

As already mentioned in the previous subsection, on the time scale much longer than Rouse time (i.e. for $t \gg \tau_R$) the sum in \eq{displacement} is dominated by the zero-th term which keeps growing indefinitely while all the other terms saturate. In this regime the monomer, together with the rest of the chain, undergoes simple diffusion with diffusion coefficient equal to $D/(N+1)$, where $D$ is the coefficient for a free bead, and one expects
\be
x^2_n(t) \sim \frac{2dT}{\xi N} t, \text{ for } t \gg \tau_1 
\label{largetime_displacement}
\ee 
(compare \eq{discm}).

In the other limiting case $t \ll \tau_0$ the displacement of the bead under consideration (and of all the other beads of the chain) is much smaller than the typical bead-to-bead distance $a$, and therefore the elastic term in \eq{disrouse} (the deterministic force on the right-hand side) is essentially constant. In this case the bead motion can be described as that of a Brownian particle dragged by a constant force. Since at the short timescales the random force always dominates the constant force (compare \eq{normal_short-time}) one expects 
\be
x^2_n(t) \sim t, \text{ for } t \ll \tau_0. 
\ee 

The most interesting behavior, however, is observed at the intermediate timescales $\tau_1 \gg t \gg \tau_0$, when beads undergo a peculiar subdiffusive dynamics which is the main characteristic feature of the Rouse model. In this regime, part of the Rouse modes are already thermalized, and part are still relaxing, with the number of thermalized modes growing with time. The exact value of $x^2_n(t)$ in this regime does depend on $n$ via the numerical values of the constants $\alpha_p^{(n)}$. However, since all of these constants are fixed numbers of the same order $N^{-1/2}$, one naturally expects that the scaling behavior of $x^2$ is $n$-independent. It is most convenient to extract it by considering a variable which has a simple expression in terms of normal modes. As an example of such a convenient variable let us choose the polymer end-to-end distance $\ve R(t) = \ve r_N - \ve r_0$, and introduce
\be
X^2(t) = \langle (\ve R(t) - \ve R(0))^2\rangle \approx \frac{16}{N}\sum_{p=1,3,5...} \langle (\ve u_p(t) - \ve u_p(0))^2.
\label{end-to-end}
\ee
Using \eq{dispu} for the dynamics of the normal modes, and the fact that for $t \gg \tau_0$ one can replace the discrete dispersion relation \eq{distimes} with a continuous one \eq{contimes}, we obtain
\be
\begin{array}{rll} 
X^2(t) &=&\displaystyle \frac{16 d T}{N \xi} \sum_{p=1,3,5...}\tau_p\left(1 - \exp\left(-\frac{2t}{\tau_p}\right)\right) \medskip \\
&=& \displaystyle \frac{16 R^2}{\pi^2}\sum_{p=1,3,5...}\frac{1}{p^2}\left(1 - \exp\left(-\frac{2tp^2}{\tau_R}\right)\right),
\end{array}
\label{ev}
\ee
where we used \eq{rouse_time} for the Rouse time and \eq{r_ideal} for the equilibrium end-to-end distance $R$ of an ideal coil. Now, if $t \ll \tau_R$ there are many terms in \eq{ev} contributing to the sum and the sum can be approximated by an integral. Therefore, in this limit the dimensional analysis immediately gives 
\be
\begin{array}{rll} 
X^2(t) &\sim & \displaystyle R^2 \int_1^{\infty}\left(1 - \exp\left(-\frac{2tp^2}{\tau_R}\right)\right) \frac{dp}{p^2} \medskip \\
& \sim & \displaystyle R^2 \tau_R^{-1/2} t^{1/2} \int_0^{\infty} y^{-2} (1-\exp(-y^2)) dy \sim \sqrt{\frac{T a^2}{\xi}} \sqrt{t},
\end{array}
\label{ev}
\ee
where the integral in the second line converges to some constant of order 1. Note that the final result does not depend on the total length of the chain, which is to be expected because the largest relaxation time of the chain, $\tau_R$, can be interpreted as a time at which information can pass from one end of the chain to another, and at which two chain ends start ``feeling'' each other. Therefore, at $t \ll \tau_R$ the ends are moving independently, and each of them is moving as a part of a semi-infinite chain. 

Thus, at intermediate times the relaxation of end-to-end distance is described by a Gaussian subdiffusive process whose dispersion grows as $ X^2(t) \sim t^{\alpha}$ with $\alpha = 1/2$. If one starts from equilibrium initial conditions (which corresponds in this particular case to end-to-end vector being normally distributed with mean-square value of \eq{r_ideal}), then this process is also obviously stationary in time. Such a process is known to be unique and is called fractal Brownian motion in the literature \cite{klafter_metzler, metzler_review}. 

The same behavior is typical of monomer coordinate $x^2_n(t)$ at intermediate times. Indeed, instead of \eq{end-to-end} one gets something like 
\be
x_n^2(t) = \frac{1}{N} \langle \left(\textbf u_0(t)-\textbf u_0(0)\right)^2\rangle +\frac{4}{N}\sum_{p=1}^{N} \cos^2\frac{p\pi n}{N} \langle \left(\textbf u_p(t)-\textbf u_p(0)\right)^2\rangle.
\ee
Since for $\tau_R \gg t \gg \tau_0$ the contribution of the first term is negligible and $\cos^2 (p\pi n/N)$ is a function rapidly oscillating between zero and one, the resulting asymptotic behavior is the same as \eq{ev} (although numerical coefficients are, of course, different).

In conclusion, let us consider the relaxation behavior of a chain fragment of a given length $s$ in a spirit similar to the discussion of $R(s)$ for different polymer states in section 1. Let us introduce a vector connecting two beads at contour distance $s$, $\ve R_{n,n+s}(t) = \ve r_{n+s} - \ve r_n$, and the corresponding time-dependent dispersion
\be
X^2(s,t) = \langle (\ve R_{n,n+s}(t) - \ve R_{n,n+s}(0))^2\rangle,
\ee
where averaging is taken both over the bead position $n$ and over realizations of thermal disorder. The chain fragment under discussion is only weakly connected to the rest of the chain by two elastic forces at $(n+s)$-th and $(n-1)$-th links. As a result, if $s \gg 1$ the fragment thermalizes almost in the same way as a free chain of the same length, its maximal relaxation time being:
\be
\tau_{\max}(s) \approx \tau_R(N=s) = s^2 \frac{\xi a^2}{\pi^2 dT}.
\ee
Now, the behavior of $X^2(s,t)$ is, up to numerical constants, the same as the behavior of $X^2(t)$ for a chain of length $N=s$. Thus, for $\tau_{\max}(s)> t >\tau_0$, $X^2(s,t)$ exhibits a subdiffusive behavior similar to \eq{ev}. In turn, for $t > \tau_{\max}(s)$ the whole fragment of length $s$ moves collectively and $X^2(s,t)$ saturates to its long-time limit of $2 R^2 (s)$. 

\subsubsection{ Scaling interpretation of the Rouse model}

The Rouse model described and solved in this section is a rare example of an exactly solvable model in polymer dynamics. However, as we have underlined in the beginning of this section, it is an oversimplified model which describes the movement of an ideal phantom chain in the absence of hydrodynamic interactions. As a result, in many polymer systems the predictions of Rouse model do not hold even qualitatively and are not directly applicable\footnote{Except for dilute solutions of relatively short chains in very viscous solvents, and, most importantly, melts and semi-dilute solutions of chains with $N<N_e$, see section 3 for more details.}. However, the Rouse model provides some important insights into how polymer chains move, and most of these insights stay valid in other, more complicated models of polymer dynamics. Let us outline them here.

If one considers thermal movement of two monomer units of a chain, these monomer units at first look like independently moving particles. However, if we wait long enough we notice that these two monomer units cannot go too far away from each other and therefore are moving in a coherent manner. The typical crossover time from independent to coherent movement, which we have called above $\tau_{\max}$, depends on the contour distance between the monomer units $s$, and is a monotonically increasing function of $s$. 

The crossover time $\tau_{\max}(N)$ corresponding to the contour distance equal to the length of the whole chain, defines the maximal relaxation time of the chain: at time scales larger than $\tau_{\max}(N)$ the chain moves essentially as a single object (colloid particle). Therefore, in a Newtonian fluid it undergoes normal diffusion with some effective diffusion coefficient. At times smaller than $\tau_{\max}(N)$ a chain can be separated into parts ({\it dynamical blobs}) such that chain fragments inside the same blob are moving coherently, while the dynamics of different blobs is independent from each other. The spatial size of such a blob $g(t)$ is of order of the typical monomer displacement at time $t$, otherwise it would not be possible for different blobs to move independently. Thus, $g(t)$ and the corresponding contour length $s_b(t)$ satisfy
\be
g(t) \approx x(t), \;\; \tau_{\max}(s_b(t)) = t.
\label{dynamic_blob}
\ee

Note that these basic insights, combined with the assumptions that chain conformation is ideal and that thermal forces at different monomer units are statistically independent, allow to rederive the main qualitative results of the Rouse model, such as \eq{largetime_displacement} and \eq{ev}\cite{tamm15}. Indeed, the independence of thermal forces means that the effective diffusion coefficient of a coherently moving domain is inversely proportional to the number of monomer units in it. At long times it immediately leads to \eq{largetime_displacement}, while at intermediate times it means that
\be
x^2(t) \sim D_{\text{eff}} t \sim \frac{D_0}{s_b(t)} t,
\label{d_eff}
\ee
where $D_0$ is a single bead diffusion coefficient. 

In turn, the fact that we are considering ideal chain dictates 
\be
g(t) \sim \sqrt{s_b(t)} a
\label{blobsize_ideal}
\ee
(compare \eq{r_ideal}). Substituting these two formulae into \eq{dynamic_blob} gives 
\be
s_b(t) \sim \sqrt{\frac{D_0 t}{a}}; \;\; \tau_{\max} (s_b) \sim \frac{a}{D_0} s_b^2 = \tau_0 s_b^2: \;\; x^2(t) \sim a^2 \sqrt{\frac{t}{\tau_0}},
\label{scaling_rouse}
\ee
where we reintroduce the typical microscopic relaxation time $\tau_0 \sim a/D_0$. Comparing \eq{scaling_rouse} to \eq{microtime} and \eq{ev} one sees that we have thus obtained, up to numerical constants, the correct intermediate asymptotic of the monomer movement. When trying to generalize Rouse model to systems with volume and topological interactions it is instructive to keep in mind this scaling derivation of Rouse results.

\subsection{Volume and topological interactions. Scaling approach and beta model}

\subsubsection{Scaling theory for swollen coils and crumpled globules}

As discussed above, the Rouse model assumes that monomer units of the chain (or beads in the beads-and-springs model) have only nearest-neighbor interactions along the chain. We know from our discussion of equilibrium properties in section 1 that this is usually not the case. Monomers typically interact with each other via forces that depend on their distance in space, regardless of whether they are adjacent along the chain or not (volume interactions). These forces can be easily added to Rouse equations \eq{disrouse} by replacing the potential energy in \eq{langevin_1} with a potential allowing for volume interactions, $U_{vol}$ in terms of \eq{part_volume}. However, as a result all the equations in such a modified Rouse model become coupled with each other and there is no hope of solving such a system of equations exactly. 

One can, nevertheless, hope to make some progress, at least for the case of swollen polymer coils, along the way of scaling arguments introduced in section 2.1.4. Indeed, as discussed in section 1.3, there is an important similarity between ideal and swollen polymer coils: both are fractal objects, i.e. end-to-end distance (and gyration radius) of a subchain of length $s$ is a power-law function of $s$
\be
R(s) = a s^{1/d_f}
\ee
with fractal dimension $d_f$ equal to 2 for ideal coils, and $d_f = 1/\nu \approx (d+2)/d$ for swollen coils. Therefore, one can expect that reasoning presented in section 2.1.4 can be applied to the swollen coils as well. The only change will be in the connection between the size of a dynamical blob and the number of monomers in it, \eq{blobsize_ideal} should be replaced with
\be
g(t) \sim (s_b(t))^{\nu} a = (s_b(t))^{1/d_f} a.
\label{blobsize_swollen}
\ee
This equation gives rise to a generalized Rouse model which gives the following predictions for the swollen coils. First, the diffusion coefficient of the whole chain is still inversely proportional to the number of monomer units. Second, the maximal relaxation time of a chain fragment of length $s$ is 
\be
\tau_{\max}(s) \sim s^{(d_f+2)/d_f} \tau_N
\label{tau_swollen}
\ee
and the maximal relaxation time for the whole chain (analogue of the Rouse time for ideal chains) scales as $\tau_{\max}(N) \sim N{(d_f+2)/d_f} \sim N^{11/5}$ for swollen coils in 3d. Third, at intermediate times much larger than the microscopic time $\tau_N$ and much smaller than this maximal relaxation time,
\be
x^2(t) \sim a^2 \left(\frac{t}{\tau_N}\right)^{2/(d_f+2)} \sim a^2 \left(\frac{t}{\tau_N}\right)^{6/11},
\label{scaling_displacement}
\ee
where the last expression corresponds to the swollen state in 3d. Importantly, the movement of a swollen chain is, generally speaking, not a Gaussian process anymore, so this estimate of the second moment, while being important, in not enough to define the process completely.

The idea of such a generalization of the Rouse model apparently goes back to de Gennes \cite{deGennes_swollen}. It is quite widely used\cite{kardar_translocation1,kardar_translocation2} and even sometimes mentioned in polymer textbooks\cite{rubinstein}. Recently we suggested\cite{tamm15} using similar reasoning to describe the dynamics of topologically controlled states of polymer chains (crumpled globule, compactized rings, etc.). Indeed, it is reasonably well established that on the large scales polymer conformations in these states are fractal with fractal dimension $d_f = 3$. This allows us to use the same reasoning as above to obtain the predictions for 
\be
x^2(t) \sim t^{2/5}; \;\; \tau_{\max}(s) \sim s^{5/3},
\label{crumpled_scaling}
\ee
the former of which seem to be in good agreement with the experimental data on chromosome dynamics \cite{garini1,garini2,garini3,lagomarsino1,lagomarsino2}, as well as some numerical results\cite{tamm15,jost_private}.

 Let us note that this approach is not free from controversy. Indeed, for ideal and swollen coils in dilute solutions it is well established that non-phantomness of the chains, which is neglected by the Rouse model, does not influence the dynamics in any substantial way. But properties of topological states are direct consequences of this non-phantomness. The assumption that one can use a generalized Rouse model for the description of the dynamics of these states is based on a hypothesis that the only effect produced by non-phantomness is the change in the fractal dimensionality of the chain packing, and that topological interactions do not influence the chain dynamics in any way other than preserving the constant fractal dimension of 3. This hypothesis, although plausible at least in some cases, is somewhat arbitrary, and the topic is far from being settled (for alternative theories of compactized ring dynamics see \cite{smrek_dynamics,ge16}). We will address this topic further in section 3.

\subsubsection{Beta model}

\begin{figure}[!t] \centering
\includegraphics[width=0.5\textwidth]{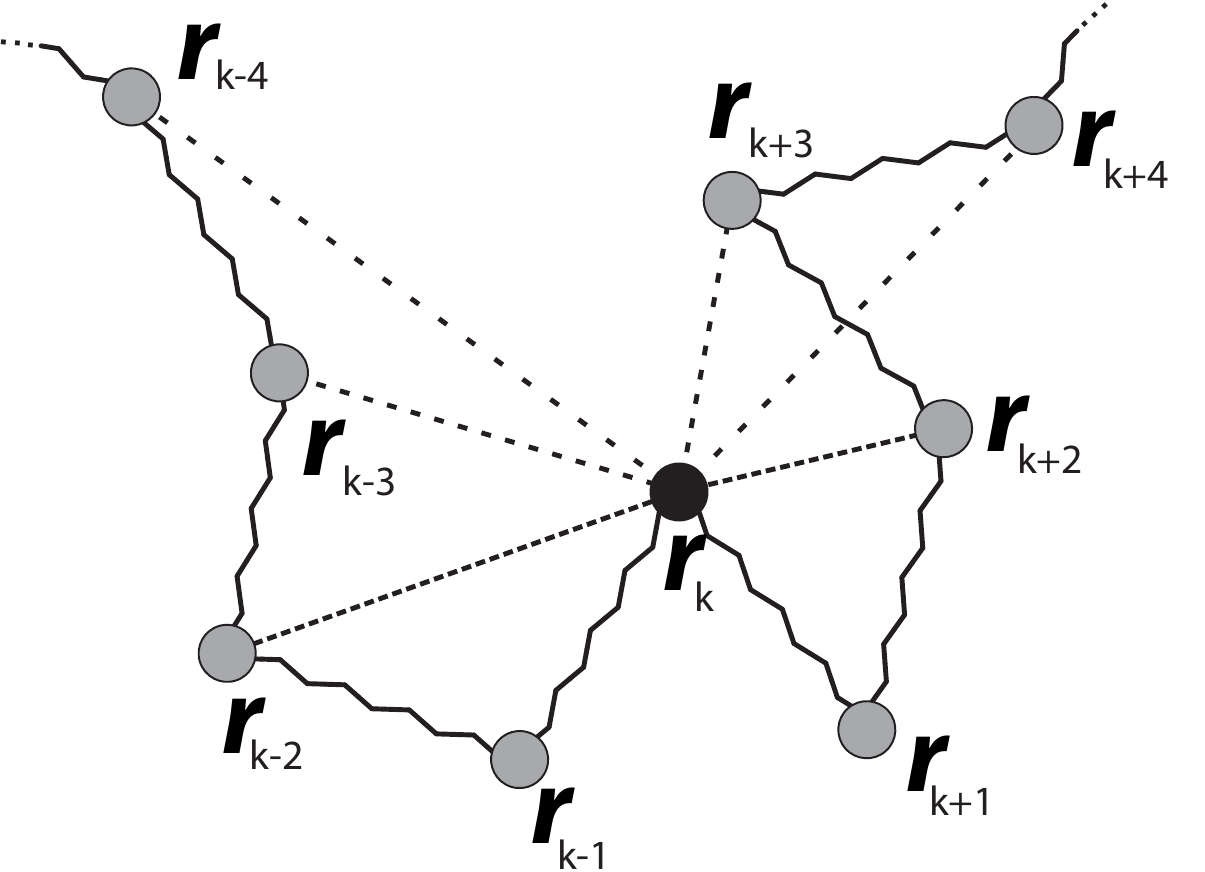}
\caption{Schematic representation of the Gaussian network model \eq{beta_potential}: each bead is connected by Gaussian springs not only with its nearest neighbors, but also with all the other beads (these new springs are shown with dashed lines. Figure adopted from \cite{polovnikov17}.}
\label{fig:4a}
\end{figure}

Scaling considerations presented above give many important insights into the dynamics of non-ideal polymer conformations in a Rouse-like regime. However, of course, they do not give a complete description of the system, even for times between $\tau_N$ and $\tau_{\max}$ where it is valid. To give the reader just one example, consider the two-point correlation function $G(s,t)$ for two monomers of the chain defined as
\be
G(s,t|N) = \overline{\langle(\ve r_n(t) - \ve r_{n+s} (t))(\ve r_n(0) - \ve r_{n+s} (0)) \rangle}.
\ee
For $t=0$ this function equals $R(s)$ and with growing $t$ it decays to zero. Now, the scaling theory presented above dictates\cite{tamm17} that this function has a universal scaling form 
\be
G(s,t|N) = A s^{2/d_f} {\cal G} \left(t s^{-(2+d_f)/d_f}|s/N\right)
\label{scaling_g}
\ee
but it says nothing at all about the possible shape and properties of the scaling function ${\cal G} (x| \eps)$. It is natural therefore to go beyond scaling theory and try to construct some analytically tractable generalization of the Rouse model. Unfortunately, as mentioned above, the most natural way of doing it -- namely, plugging the explicit microscopically justified form of volume interactions into the Rouse equations -- leads to equations intractable analytically. 
A possible way to try and circumvent this problem is to construct an artificial analytically tractable Hamiltonian, which would, although wrong on the microscopic level, mimic some of the large-scale properties of the system. As an example one can try and replace the elastic energy \eq{energy_ideal} with a similar but more general potential of a Gaussian network\cite{bahar97,haliloglu97}
\be
U(\ve r_0, ..., \ve r_N) = \frac{1}{2} \sum_{n>m}k_{nm} (\ve r_n - \ve r_m)^2 = \frac{1}{2}\sum_{n=0}^{n=N} \ve r_n^2 \sum_{m \ne n} k_{ij} - \sum_{n>m} k_{nm} \ve r_n \ve r_m,
\label{beta_potential}
\ee
leading to a set of Langevin equations
\be
\xi \frac{d \ve r_n}{dt} = \sum_{m \ne n}k_{nm} (\ve r_m - \ve r_n) + \ve F_n,\;\;\; n=0,...,N.
\label{langevin_2}
\ee
This means essentially that instead of connecting beads adjacent along the chain with elastic springs, we now introduce springs between all pairs of beads, with spring stiffness $k_{nm}$ being a function of bead positions $n$ and $m$ (see \fig{fig:4a}). The original bead-and-springs model is, of course, recovered if stiffness matrix is tridiagonal, $k_{nm} = (d T/a^2) \delta_{|n-m|,1}$. For various applications related to protein folding and other heteropolymer problems it makes sense to make $k_{nm}$ a non-trivial function of both positions. Here we limit ourselves to homopolymer chains which have a property of translational invariance along the chain, and therefore $k_{nm}$ should only be a function of $s = |n-m|$. 

It is easy to show that if $k_{n,n+s}$ is a rapidly (faster than power law) decaying function of $s$, the resulting Langevin dynamics will not, on the large scale, differ from Rouse model. Qualitatively, it follows from simple renormalization group considerations. As we have outlined in section 1.1, the choice of the size of a bead is somewhat arbitrary and is only bounded from below. Therefore, large-scale properties of polymer chains are universal with respect to a renormalization-group-like transformation: unite several adjacent beads into one, and, with a proper renormalization of bead-to-bead distance and interaction parameters, the resulting system will have exactly the same large-scale properties. If the interaction constants $k_{n,n+s}$ decay fast enough then under such a renormalization all the interactions except for the nearest-neighbor ones will converge to zero, and therefore the system will converge to the original Rouse model in large-scale properties. 

Behavior of systems with $k_{n,n+s}$ decaying as a power law of $s$ is much richer (see recent ref. \cite{polovnikov17} for a detailed analysis). Here we will discuss one particular way of introducing such a power-law-decaying stiffness $k$ suggested recently  under the name of {\it beta model}\cite{Amitai2013}. In our presentation we mostly follow \cite{tamm17}. Beta model corresponds to a particular choice of $k_{nm}$ such that in standard Rouse coordinates \eq{Fourier} the generalized potential takes the form similar to \eq{normal_potential}:
\be
U (\ve u_0, ..., \ve u_N) = \frac{1}{2}\sum_{p=1}^{N} \tilde {\kappa}_p \ve u_p^2
\label{beta_energy}
\ee
but with modified spring constants
\be
\tilde {\kappa}_p = \frac{4 d T}{a^2}\sin^{\beta}\left( \frac{\pi}{2}\frac{p}{N}\right)
\ee
that dependent on a free parameter $\beta$ (where $\beta = 2$ corresponds to the regular Rouse model).

Using the inverse Fourier transform, one gets the following result for the original parameters of the potential $k_{nm}$:
\be
U (\ve r_0, ..., \ve r_N) = \frac{1}{2} \sum_{p=1}^{N} \tilde{\kappa}_p \left(\sum_{n=0}^{N} \ve r_n \alpha_p^{(n)} \right)^2 =
\frac{1}{2} \sum_{n,m = 0}^N A_{nm} \ve r_n \ve r_m,
\label{newA}
\ee
where
\be
\begin{array}{rll}
A_{nm} &=& \displaystyle \sum_{p=1}^{N} \tilde{\kappa}_p \alpha_p^{(n)} \alpha_p^{(m)} = \medskip \\
&=&\displaystyle \frac{8d T}{N a^2} \sum_{p=1}^{N} \sin^\beta \left(\frac{p\pi}{2N}\right)
\cos\left(\frac{\pi p(n-1/2)}{N}\right)\cos\left(\frac{\pi p(m-1/2)}{N}\right)
\end{array}
\label{inverse}
\ee
Comparing \eq{beta_potential} and \eq{newA} gives $A_{nn} = \sum_{m \ne n} k_{nm}$ and $A_{nm} = -2 k_{nm}$ for $n \ne m$. For the special case of $\beta =2$, $A_{nm} = 2(\delta_{n-m, 0} - \delta_{|n-m|, 1})$ and we recover the Rouse model.

Similarly to the Rouse model, in normal coordinates the relaxation of the beta-model can be understood in terms of a set of simultaneously relaxing Ornstein-Uhlenbeck oscillators with relaxation times $\tilde{\tau}_p = \xi \tilde{\kappa}_p^{-1}$. The continuous limit of the model corresponds to taking $p\ll N$ and expressing all times in terms of the maximal relaxation time 
\be
\tau_{\beta} =\tau_1 =\frac{\xi a^2}{4dT} \sin^{-\beta}\left(\frac{\pi}{2N}\right) \approx \frac{2^{\beta-2}}{\pi^{\beta}d} \frac{\xi a^2}{T} N^{\beta}. 
\label{tau_beta}
\ee
In this limit the relaxation time of the $p$-th mode of the beta model is simply $\tilde{\tau}_p = \tau_{\beta} p^{-\beta} $.

\subsubsection{Equilibrium conformations in beta model}

So far, the beta model is introduced in a somewhat formal way. To make some intuition about what equations \eq{beta_potential},\eq{langevin_2}, and \eq{beta_energy} actually imply, let us estimate the equilibrium distance $R(s)$ between monomer units separated by contour distance $s$: \be
\begin{array}{rll}
R^2(s) &=& \displaystyle \overline{(\ve r_n - \ve r_{n+s})^2} = \sum_{p=1}^{N} \overline{\ve u_p^2} (\alpha_p^{(n)}-\alpha_p^{(n+s)})^2 \medskip \\
& =& \displaystyle \sum_{p=1}^{N} \frac{a^2}{4} \sin^{-\beta}\left( \frac{\pi p}{2N}\right) (\alpha_p^{(n)}-\alpha_p^{(n+s)})^2
\end{array} 
\ee
where we took into account that $\alpha_0^{(n)}$ is independent of $n$, and that the average values of normal coordinates at equilibrium $\overline{\ve u_p^2}$ are defined by the equipartition theorem. Substituting \eq{alphas} and averaging the right hand side of this expression over $n$, one gets, up to numerical constants, the following estimate:
\be
\begin{array}{rll}
R^2(s)& \sim& \displaystyle \frac{a^2}{N}\sum_{p=1}^{N} \sin^{-\beta}\left( \frac{\pi p}{2N}\right) \sin^2 \left( \frac{\pi p s}{2N}\right) \medskip \\
& \sim & \displaystyle a^2 \int_0^{1/2} \sin^{-\beta}(\pi z) \sin^2(\pi s z) dz.
\end{array}
\ee
Due to the singularity in the first multiplier, the main contribution into the integral comes from the vicinity of $z=0$ where the first sine can be replaced by its argument. As a result, after replacing $ s z \to y $ one gets
\be
R^2(s) \sim a^2 s^{\beta -1} \times (\text{some integral independent of }s)
\ee
Therefore, the equilibrium conformations of polymer chains with beta-model interaction potential \eq{beta_energy} are fractal (i.e., $R(s)$ is a power law function of $s$) with fractal dimension 
\be
d_f = 2/(\beta-1).
\label{beta_dimension}
\ee 
This means, for example, that $\beta = 11/5$ emulates the swollen coil in 3d, $\beta = 5/2$ stands for the swollen coil in 2d, and $\beta = 5/3$ for compact topological states of rings and rapidly collapsed chains.

\subsubsection{Single monomer displacement in beta model}

Consider now the dynamics of a single monomer in beta model. The mean-square monomer displacement $x^2(t)$ is given by
\be
x_n^2(t) = \sum_{p=1}^N \overline{\langle (\ve u_p(t) - \ve u_p(0))^2} \left(\alpha_p^{(n)}\right)^2 \sim \frac{1}{N}\sum_{p=1}^N \overline{\ve u_p^2} \left(1 - \exp\left(-\frac{2t}{\tilde{\tau}_p}\right)\right).
\ee 
After replacing the sum with an integral and using the dispersion relation $\tilde{\tau}_p = \tau_{\beta} p^{-\beta}$, which is valid for $p \ll N$, this leads to
\be
x_n^2(t) \sim \frac{d T}{N \xi} \tau_{\beta} \int_1^{\infty} p^{-\beta} \left(1 - \exp\left(-\frac{2t p^{\beta}}{\tau_{\beta}}\right)\right) dp \sim N^{\beta-1} a^2 \left(\frac{t}{\tau_{\beta}}\right)^{1-1/\beta}, 
\label{beta_displacement}
\ee
where we used \eq{tau_beta} and the last estimate is valid for $t \ll \tau_{\beta}$. This result is to be compared to the corresponding result for the Rouse model \eq{ev} and the scaling prediction \eq{scaling_displacement}. Using the connection \eq{beta_dimension} between the model parameter $\beta$ and the equilibrium fractal dimension of the structure $d_f$, one can check immediately that equations \eq{scaling_displacement} and \eq{beta_displacement} coincide. Note that, as in the case of Rouse model, the last equation can, with the use of \eq{tau_beta}, be rewritten in the form
\be
x_n^2(t) \sim a^2 \left(\frac {t T}{\xi a^2}\right) ^{1-1/\beta} \sim a^2 \left(\frac{t}{\tau_0}\right)^{1-1/\beta} 
\label{beta_displacement_2}
\ee 

making the result independent of the full length of the chain $N$.

\subsubsection {Monomer-monomer correlations in the beta model}

Most importantly, the beta model allows to go beyond rederivation of the scaling exponents and to obtain some insight into how scaling functions actually look like. As an example, we provide here a brief summary of recent results\cite{tamm17} concerning the scaling function ${\cal G} (x| \eps)$ defined in \eq{scaling_g}. This function describes how autocorrelations of the radius vector $(\ve r_{n+s} - \ve r_n)$ decay with renormalized time $x = t s^{-\beta}/\tau_N$, and, generally speaking, it depends on $\eps = s/N$ as a parameter. 

This scaling function can be expressed, up to numerical parameters, in terms of the following series
\begin{equation}
{\cal G}(x|\eps) \sim
\sum_{k=1}^{\infty} (-1)^{k+1}
\frac{\pi^{2k}}{(2k)!} \int_{\eps}^{1} y^{2k-\beta} \exp
\left( - 2 x y^{\beta}\right) dy,
\label{corr_function_master}
\end{equation} 
Importantly, this expression is valid not only for $\eps = 0$ but also for some small positive $\eps$, allowing to consider long but finite chains, while the scaling theory is only exact for infinite chains. Analyzing the asymptotics of this series one obtains following three regimes in the $\eps \ll 1$ limit.

At the short-time limit $x \ll 1$ one gets
\begin{equation}
{\cal G}(x|\eps) = 1 - \exp \left(- C x \right)
\label{short_time}
\end{equation}
with some numerical parameter $C$, independently of $\eps$. At intermediate times, $\epsilon^{-\beta}\gg x \gg 1$ (i.e., at the time-scale much smaller than the relaxation time of the whole chain) the correlations decay as a power law for the physically interesting case of $3/2 \leq \beta < 3$ 
\begin{equation}
{\cal G}(x|\eps) \sim x^{(\beta-3)/\beta}
\label{intermediate_time}
\end{equation}
and is once again independent of $\eps$. This intermediate regime is exactly the regime of validity of the scaling law for the single monomer subdiffusion in \eq{beta_displacement_2}. Finally, for small but finite $\eps$ the third regime, $x \gg \eps^{-\beta}$ corresponds to collective diffusion of the whole chain. In this regime correlations decay exponentially which is typical for simple diffusion:
\begin{equation}
{\cal G}(x|\eps) \sim \eps^{1 - 2\beta } \exp \left(- C x \eps^{\beta}\right).
\label{long_time}
\end{equation}
Note that the expression under exponent 
\be
x \eps^{\beta} = \frac{ t N^{\beta}} {\tau_N} \sim \frac{t}{\tau_{\beta}} 
\ee
is independent of $s$ and depends only on the relaxation time of the whole chain $\tau_{\beta}$.

\subsection{ Rouse model in viscoelastic medium}

Recently, there have been another interesting development in the study of the dynamics of unentangled polymer systems, which we should mention very briefly here. Throughout previous subsections we assumed that the solvent surrounding the polymer chain is a simple Newtonian liquid, so that the friction acting on the beads constituting polymer chain is inversely proportional to their velocity. Although such behavior of the solvent is, of course, most common, it is, generally speaking not always the case. It is known empirically that some liquids, in particular biological ones, can behave in a more complicated viscoelastic way with friction acting on the probe particle depending not only on the instantaneous velocity, but on the history of motion as well. In recent years there have been several works discussing the behavior of Rouse\cite{theriot10, lampo16} and beta models\cite{tamm17} in such non-Newtonian solvents. 

The most mathematically simple way of describing such a solvent with history-dependent friction is to write a generalized Langevin equation of motion for a single probe particle in the form 
\begin{equation}
\xi_{\alpha} \int_0^t dt' K_{\alpha}(t-t') \frac{d \ve r(t')}{dt'} = \ve F_{\alpha} (t),
\label{fractal_langevin}
\end{equation}
with a powel-law kernel 
\begin{equation}
K_{\alpha}(t) = \frac{(2-\alpha)(1-\alpha)}{|t|^\alpha}
\label{kernel}
\end{equation}
with $\alpha = 1$ corresponding to the usual Newtonian case. Fluctuation-dissipation theorem then dictates that the thermal noise in the right hand side of \eq{fractal_langevin} is no longer delta-correlated but satisfies
\be
\langle \ve F_{\alpha}(t) \ve F_{\alpha}(t') \rangle \sim 2d T \xi_{\alpha} K_{\alpha}(t-t')
\ee

The simplest way of generalizing Rouse or beta model to the case of such a viscoelastic solvent is to replace friction and random force in Langevin equations \eq{langevin_2} while leaving the potentials of bead-to-bead interactions the same, and assuming that random forces acting on different beads are still independent\footnote{Note that this last assumption is somewhat arbitrary: it is natural to assume that in a complex liquid with memory forces acting on different beads might be correlated. The exact nature of these correlations depends, however, on the microscopic nature of solvent viscoelasticity. Theory describing the dynamics of polymers in solvents with spatial as well as temporal memory is still, for the best of our knowledge, lacking.}. We will not present here full analysis of the resulting theory but just formulate the most important results. 

First, the viscoelasticity of the solvent does not influence the equilibrium polymer conformations. That is to say, equilibrium state of a Rouse polymer in viscoelastic solvent is still the ideal chain conformation with fractal dimension 2, while for the beta model the connection between $\beta$ and fractal dimension, \eq{beta_dimension}, remains the same. 
Second, the effects of chain connectivity and solvent viscoelasticity on the chain dynamics are multiplicative. For example, the monomer displacement at intermediate times scales as
\be
x_n^2(t) \sim a^2 \left(\frac{t}{\tau_0}\right)^{\alpha(1-1/\beta)} 
\label{beta_displacement_viscoelastic}
\ee 
with $\beta = 2$ corresponding to Rouse model, and $\alpha = 1$ -- to the Newtonian solvent. Third, it turns out that it is possible to get quite detailed results for the two-point correlation function in this model, akin to those presented in section 2.2.5. The detailed discussion of this results goes beyond the scope of the current chapter, we address the reader to original papers\cite{lampo16, tamm17} for full details.

\subsection{ Hydrodynamic interactions and Zimm model}

\subsubsection{Hydrodynamic interactions: a qualitative discussion}

In the previous sections we discussed the dynamics of a single phantom polymer chain under the assumption that the chain is placed into immovable solvent which is not perturbed by the movement of the chain. Clearly, in most contexts such an assumption is not correct: while there is a friction force acting on a monomer unit, there always is, according to the law of action and reaction, a companion force acting on the surrounding fluid. This force, if the fluid is not infinitely viscous, makes the fluid move thus violating the assumption of immobility. 

\begin{figure}[!t] \centering
\includegraphics[width=0.5\textwidth]{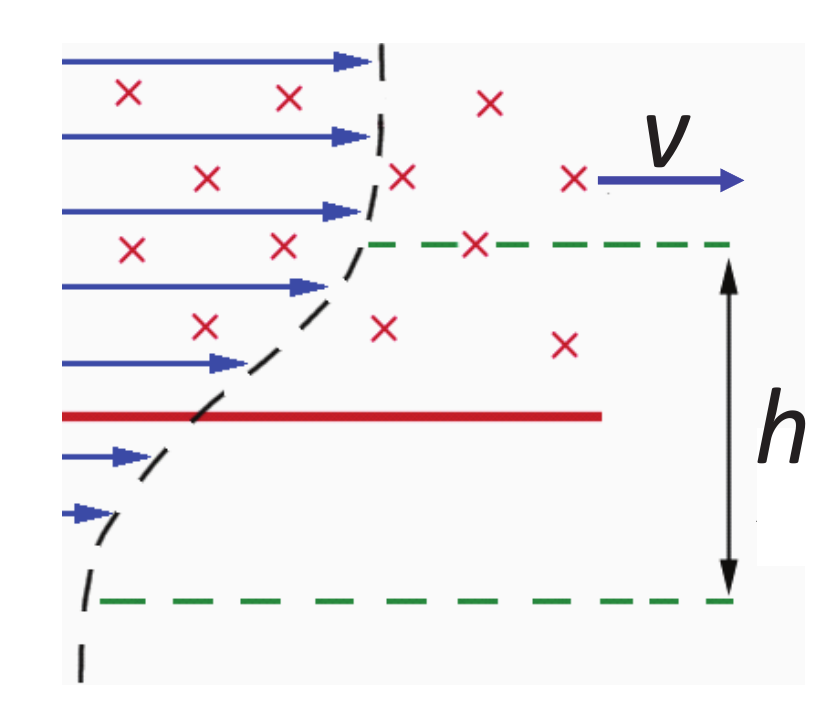}
\caption{Illustration of the auxiliary problem discussed in the text. The bottom half-space (under the red line) is filled with pure liquid. In the upper half-space, obstacles moving with constant velocity $v$ are dispersed with concentration $c$. }
\label{fig:5}
\end{figure}

How big is this effect? Is it just a small perturbation of the Rouse picture, or it can change the whole behavior of the chain? To get some insight into this question consider the following auxiliary problem (see \fig{fig:5}). Consider an infinite 3d space\footnote{Throughout this section we limit ourselves to $d=3$, hydrodynamic interactions of polymers on a two-dimensional plane are drastically different.} half of which (bottom half in \fig{fig:5}) is filled with a pure viscous liquid, and another half (top half in the figure) contains point-like obstacles with some concentration $c$. Assume now that all obstacles are moving to the right with some constant velocity $v$. In the stationary regime the velocity of the fluid $u$ will be a function of the distance $z$ to the plain separating pure solvent and solvent with moving obstacles. Clearly, $u(+\infty) = v,\, u(-\infty) = 0$, and in between there is a region of some characteristic width $h$ where the velocity changes smoothly from one limiting value to another. The velocity distribution satisfies the one-dimensional Navier-Stokes equation
\be
\eta \frac{d^2 u(z) }{dz^2} = \xi c (u(z)-v),
\ee
where $\eta$ is the viscosity of the solvent. Dimensional analysis of this equation immediately leads to an estimate
\be
h \sim \sqrt{\frac{\eta}{c \xi}}.
\ee

Now, consider an ideal polymer coil represented by the beads-and-springs model. The spatial size of this coil is given by \eq{r_ideal} and the number density of beads within a coil by \eq{density}. Assume that all beads start moving collectively with some velocity $v$. Will the coupling between them suffice to make the liquid inside the coil move together with the polymer (so-called non-draining limit), or will the typical liquid velocity be much slower than the velocity of the beads (so-called free draining)? Clearly, to understand that one should compare the size of the coil $R$ with the width of the transitional layer $h(s)$. If $R\gg h(s)$ the liquid inside the coil should move together with the polymer, while if $R \ll h(s)$ the liquid is barely perturbed by the movement of the beads. One checks immediately that for long enough chains it is the first option that occurs: if 
\be
N \gg (\eta a/ \xi)^{1/(1-\nu)}
\ee 
where $\nu =1/2$ for ideal coils, $\nu \approx 3/5$ for swollen coils, then the non-draining limit is realizied, and the solvent inside the coil moves together with it.

It is, of course, quite a significant deviation from the assumptions of the Rouse model. It means that long polymer chains in diluted solutions function in the regime dominated by hydrodynamic interactions, and we need a new modified theory to allow for this fact. In fact, Rouse model is actually much more relevant for semi-dilute and concentrated solutions where hydrodynamic interactions become screened in a way similar to what we have described in 1.5. Below we present a quantitative theory to allow for hydrodynamic interactions, the {\it Zimm model} of polymer dynamics \cite{zimm}, but before that let us make some scaling and qualitative arguments. 

\subsubsection{Scaling approach to hydrodynamic interactions}

When we discussed the role of volume interactions in polymer dynamics, it turned out to be very instructive to look into scaling concepts beyond the classical Rouse model, as described in section 2.1.4, and discuss which of these concepts hold, and which break in the new, more complicated setting. Let us try and employ this trick one more time. 

Consider the movement of the chain as a whole. First, Rouse model predicts that at the longest timescale the polymer diffuses collectively (i.e., the displacements of the monomer units relative to each other become negligible compared to the displacement of the center of mass). Clearly, this statement should not change. Second, in Rouse model all friction forces between the monomers and the solvent are independent, and therefore the collective diffusion coefficient is inversely proportional to the number of beads. This is no longer true in the presence of hydrodynamic interactions. However, in the non-draining limit it is very easy to get the collective diffusion coefficient. Indeed, the coil moves collectively and together with the trapped solvent can be seen as an impenetrable ball. The diffusion coefficient of such a ball is given by the Einstein-Stokes relation:
\be
D \sim\frac{T}{\eta R} \sim \frac{T}{\eta a N^{\nu}},
\label{d}
\ee
the result which is in a very good agreement with experimental data on the diffusivity in diluted solutions of long polymer chains.

Now, at times smaller than the relaxation time of the whole chain $\tau_{\max}$ (which in this case is still to be determined), the blob picture presented in section 2.1.4 still holds. The only difference is the diffusion coefficient of the blob, which is now inversely proportional to the spatial size of the blob, not the number of monomer units in it. Substituting 
\be
D_{\text{eff}} = \frac{D_0}{g/a} 
\ee
into \eq{dynamic_blob}, \eq{d_eff}, and \eq{blobsize_ideal} gives the following results for the blob relaxation time and monomer displacement:
\be
x^2 (t) \sim a^2 \left(\frac{t}{\tau_N} \right)^{2/3}; \; \; \tau_{\max} (s) \sim \tau_N s^{3 \nu} 
\label{zimm_scaling}
\ee
and thus the maximal relaxation time of the whole chain, the so-called {\it Zimm time} is
\be
\tau_{\max}(N) = \tau_Z = \tau_N N^{3\nu}.
\ee
These scaling arguments are valid in the non-draining limit for both ideal and swollen coils. Let us now consider a more quantitative picture of the monomer dynamics in the more simple ideal case.

\subsubsection{Quantitative Zimm model}

We will show below that the dynamics of an ideal chain in the presence of hydrodynamic interactions can be approximately described by a set of Rouse-like equations on the normal coordinates $\ve u_p$, but with renormalized $p$-dependent friction coefficients $\xi_p$. 

To develop an analytical treatment of the Zimm model, one has to change the Langevin equations describing the movements of the beads, taking into account the movement of the solvent. This means that the friction force acting on a bead will be modified as follows:
\be
\textbf F_n^{\text{fr}} = -\xi\left(\frac{d \ve r_n}{d t} - \textbf v(\textbf r_n)\right),
\label{zimm_friction}
\ee
where $\textbf v(\textbf r_n)$ is the value of solvent velocity field at $\ve r_N$. In turn, the dynamics of solvent (incompressible and moving with small Reynolds numbers), is described by the Navier-Stokes equation
\be
\eta \Delta \textbf v(\ve r) = \nabla p (\ve r) - \ve f(\ve r),
\label{navier}
\ee
(where $p(\textbf r)$ is pressure and $\ve f (\textbf r)$ is the volume density of external force), and the incompressibility condition
\be
\nabla \ve v (\ve r) = 0.
\ee
In Fourier space (under transition $\ve X_k = \int \textbf X(\textbf r) \exp(i\textbf k \textbf r) d \textbf r)$ these equations take the form
\be
-\eta k^2 \textbf v_k -i\textbf k p_k + \ve f_k = 0; \qquad \textbf k \textbf v_k = 0,
\label{navier-f}
\ee
which leads, excluding the pressure, to:
\be
\textbf v_k = \frac{\ve f_k - \ve n_k (\ve n_k \ve f_k)}{\eta k^2}, 
\label{vk}
\ee
where we introduced the unit vector in the direction of $\ve k$, $\textbf n_{\textbf k} = \frac{\textbf k}{|\textbf k|}$. Making the inverse Fourier transform of \eq{vk} gives:
\be
\textbf v(\textbf r) = \int \hat{\ve H}(\textbf r - \textbf r') \ve f(\textbf r') d \textbf r',
\label{vh}
\ee
where the operator $\hat{\ve H}$ is called the Oseen tensor and its components read
\be
H_{\alpha \beta}(\textbf r) = \frac{1}{8\pi\eta r}\left(\delta_{\alpha \beta} + (\textbf n_{\textbf r})_{\alpha} (\textbf n_{\textbf r})_{\beta}\right); \qquad \textbf n_{\textbf r} = \frac{\textbf r}{|\textbf r|}.
\ee

Now, due to the law of action and reaction, the force acting on the solvent is minus the friction force felt by the beads:
\be
\ve f (\textbf r) = -\sum_{n} \textbf F_n^{\text{fr}}(\textbf r_n) \delta(\textbf r - \textbf r_n).
\label{phi1}
\ee
In the overdamped regime the total force acting on a bead is zero (see \eq{langevin_1}), thus:
\be
\textbf F_n^{\text{fr}} = \frac{\partial U\left(\ve r_0,...,\ve r_N\right)}{\partial \ve r_n} - \textbf F_n.
\label{langevin_3}
\ee
Substituting these equations into \eq{vh} gives the following equation for the velocity field at the position of $n$-th bead $\ve r_n$:
\be
\textbf v(\textbf r_n) = \sum_{m \ne n} \hat{\ve H}(\textbf r_n - \textbf r_m) \left(-\frac{\partial U\left(\ve r_0,...,\ve r_N\right)}{\partial \ve r_m} + \textbf F_m \right).
\ee
In turn, substituting \eq{zimm_friction} into \eq{langevin_3}, one gets
\be
\frac{d \ve r_n}{d t} = \ve v( \ve r_n) +\frac{1}{\xi} \left(-\frac{\partial U\left(\ve r_0,...,\ve r_N\right)}{\partial \ve r_n} + \textbf F_n \right)
\ee 
Thus, if we add  $H_{\alpha\beta} (\ve r = \ve 0) = \xi^{-1} \delta_{\alpha \beta}$ to the definition of the Oseen tensor, the resulting Langevin equation becomes:
\be
\frac{d \textbf r_n}{d t} = \sum_{m} \hat{\ve H}(\textbf r_n - \textbf r_m) \left(-\frac{\partial U\left(\ve r_0,...,\ve r_N\right)}{\partial \ve r_m} + \textbf F_m \right).
\label{langh1}
\ee

The Oseen tensor $\hat{\ve H}(\textbf r)$ is non-linear in $\ve r$ making the set of equations \eq{langh1} unsovable analytically. One can, however, hope to get an approximate mean-field solution fpr the average values of displacements at equilibrium by replacing $\hat{\ve H}(\ve r_n - \ve r_m)$ with its average value at thermal equilibrium $\overline {\ve H}(s)$, $s = |n-m|$:
\be
\begin{array}{rll}
\hat{\ve H}(\ve r_n - \ve r_m) & \to& \overline {\ve H}(s);\medskip \\
\overline{H}_{\alpha\beta}(s) &=& \displaystyle \frac{1}{8\pi\eta} \overline{ |\ve r_{n,n+s}|^{-1}} \left[\overline {(\textbf n_{\textbf r})_{\alpha} (\textbf n_{\textbf r})_{\beta}} + \delta_{\alpha\beta}\right],
\end{array}
\ee
which simplifies to
\be
\overline {\ve H}(s) = \frac{\hat{\ve I}}{\sqrt{6 \pi^3 }\eta a \sqrt{s}} = H(s) \hat{\ve I}
\ee
using the equilibrium distribution of bead-to-bead distances in the ideal coil \eq{part_ideal} and the fact that $\overline {(\textbf n_{\textbf r})_{\alpha} (\textbf n_{\textbf r})_{\beta} } = \delta_{\alpha\beta}/3$. Here $\hat{\ve I}$ is the identity tensor and the last equation is the definition of scalar function $H(s)$.

Substituting the pre-averaged tensor $\overline {\ve H}(s)$ into \eq{langh1} one gets the following set of linear mean-field Zimm equations\cite{zimm}:
\be
\frac{d \textbf r_n}{d t} = \sum_{m=0}^N H(|n-m|) \left(\ve F_m^{\text{el}} + \textbf F_m \right),
\label{langh2}
\ee
where the elastic force acting on $m$-th particle in bead-and-spring model is
\be
\ve F_m^{\text{el}} = -\frac{\partial U\left(\ve r_0,...,\ve r_N\right)}{\partial \ve r_m} = \left\{
\begin{array}{ll}
k(\ve r_{m+1}+ \ve r_{m-1} -2\ve r_m & \text{ for } m = 1,...,N-1; \medskip \\
k(\ve r_1 - \ve r_0) & \text{ for } m = 0; \medskip \\
k(\ve r_{N-1} - \ve r_N) & \text{ for } m = N.
\end{array}
\right.
\ee
The right hand side of \eq{langh2} is a convolution with respect to the coordinate along the chain. Therefore, Fourier transform turns it into a product of two Fourier images, giving 
\be
\frac{d \textbf u_p(t)}{\partial t} = H_p (-\kappa_p\textbf u_p + \textbf f_p)
\ee
where $\kappa_p$ is the same as in the Rouse model and $H_p$ is the Fourier image of $H(s)$
\be
H_{p} \approx \frac{\sqrt N}{\sqrt{12 \pi^3} \eta a }p^{-1/2}
\ee
As a result, in normal coordinates the equations of the Zimm model are similar to Rouse equations \eq{OU} but for the constant friction coefficient $\xi$ being replaced with a $p$-dependent one:
\be
\begin{array}{l}
\displaystyle \xi_0 = \left[\frac{1}{N}\int_0^N dn \int_0^N dm h(|n-m|)\right]^{-1} =\sqrt{\frac{27 \pi^3}{32}} \eta a N^{-1/2} \medskip \\
\xi_p = \sqrt{12 \pi^3} \eta a N^{-1/2} \sqrt{p}, \qquad p \geq 1.
\end{array}
\ee
while the potential stiffnesses in the $p \ll N$ (continuous) limits are the same as in the Rouse model
\be
\kappa_p = \frac{3 \pi^2 T}{a^2} \left(\frac{p}{N}\right)^2 = \kappa_1 p^2
\ee
As a result, the the equation for the displacement of the center of mass \eq{discm} is still valid, and the diffusion coefficient of the chain as a whole is 
\be
D = \frac{3T}{\xi_0 N} \sim \frac{T}{\eta a N^{1/2}} \sim \frac{T}{\eta R}
\ee
as predicted by \eq{d}. The numerical coefficient which we omitted here can also be easily obtained. In turn, the relaxation times of other normal modes behave in the $p \ll N$ limit as:
\be
\tau_p = \frac {\xi_p}{\kappa_p} = \tau_1 p^{-3/2};\;\;
\tau_1 = \tau_Z= \frac{2}{\sqrt{3\pi}}\frac{ \eta a^3}{T} N^{3/2}\sim \frac{\eta R^3}{T}
\ee
once again confirming the predictions of the scaling theory above.

\subsubsection{Zimm model for swollen chains}

It is possible to generalize the mean-field approach presented in the previous session to the case of swollen polymer coils with short-range volume interactions. Indeed, the Langevin equation \eq{langh1} remains valid, at least formally, for chains with volume interactions. The only changes are {\it i} the particular form of the potential $U(\ve r_0, ... \ve r_N)$, which now depends on the distances between all pairs of beads, and {\it ii} the distribution of distances between beads. Of these two differences the latter is more important. Indeed, if the repulsive potential is very short-ranged, it can be approximated with 
\be
U(\ve r_0, ... \ve r_N) = U_{id} (\ve r_0, ... \ve r_N) + B \sum_{n\neq m} \delta(\ve r_n - \ve r_m) 
\ee
and therefore the additional force is almost always zero (except for conformations in which any two beads take the same spatial position, $\ve r_n = \ve r_m$, which happen with probability 0). As a result, one can make the same trick as in the previous section, replacing the Oseen tensor with its value averaged over the equilibrium distribution:
\be
\overline{\ve H}(s) = \displaystyle \frac{\hat{\ve I}}{6\pi\eta} \overline{ |\ve r_{n,n+s}|^{-1}} .
\ee 
Calculating this average exactly is somewhat tricky since bead-to-bead distance in the swollen coil is not normally distributed, not to mention the fact that we assume here that distances between different pairs of units can be averaged independently, which is a rather crude approximation. However, we know that the swollen coil is a fractal object with $R(s) \sim a s^{\nu}$ which allows to deduce
\be
\overline{ |\ve r_{n,n+s}|^{-1}} = \frac{A}{a s^{\nu}}
\label{hydroradius}
\ee
with some numerical constant $A$. Repeating the same analysis as presented above for ideal coils leads once again to a set of Ornstein-Uhlenbeck equations\eq{OU} for the Rouse modes with $p$-dependent friction coefficients, but with a different form of dependence on $p$ and $N$:
\be
\begin{array}{l}
\displaystyle \xi_0 \sim \eta a N^{\nu-1} \medskip \\
\xi_p \sim \xi_0 p^{1-\nu}, \qquad p \geq 1.
\end{array}
\ee
As a result we once again arrive to the scaling equations \eq{d} and \eq{zimm_scaling}.

\section {Role of entanglements in polymer dynamics}

\subsection{Role of entanglements and applicability limits of different theories}

This chapter is primarily dealing with the theories and concepts describing polymer dynamics in the case when the chain entanglements are negligible, and one can consider the chain is if it where phantom. However, our discussion would be incomplete without at least some discussion of the ways in which entanglements affect the movement of polymer chains. As a by-product of this discussion we will also discuss here the limits of applicability of the models discussed above.

To begin with, let us discuss the situation from the qualitative point of view. What the entanglement effects actually are? As an example, consider a melt of very long linear chains and try to understand how one selected chain moves in such a melt. We may notice that, since chains are impenetrable linear objects (the property which we have called non-phantomness), surrounding chains sometimes work with respect to the selected chain as something like the obstacles  in the array-of-obstacles example in section 1.6.1. For example, if one of the surrounding chains forms a tight loop or a knot around the selected chain, it will drastically reduce the possible {\it directions} of the chain movement: the chain can easily slip through the loop but can barely move in perpendicular direction. 

It is important to notice two things about this example. First, entanglement effects are a consequence of {\it other chains} being present in the vicinity of a selected chain, and therefore they should become stronger with the increase of concentration. In the dilute regime these effects are negligible: the concentration inside a single polymer coil is vanishingly small (see \eq{density}) and subchains which are far from each other along the chain vary rarely meet each other in space. As polymer concentration in solution increases, the chains start meeting each other more often and the entanglements start to appear. 

Second, not every chain-to-chain contact constitutes an entanglement. Entanglement is not just a couple of chains touching each other so that there is a friction force between them when they move, it is a substantially anisotropic restriction on chain movement. In order to form such a restriction the chains should be sufficiently long: length of the chain $N$ should exceed some critical value $N_e$, known as {\it entanglement length} in order for proper entanglements to start forming. The value of $N_e$ depends on the microscopic chemical structure of the chains, but experiments and computer simulations show that typically in polymer melts it is somewhere between 50 and 500. 

Now, in the intermediate concentration regime of semidilute solutions the length of the chains needed to form an entanglement is larger than $N_e$ for the melts, and it decreases with the increase of concentration. Let us remind that, as described in section 1.5, semidilute solutions can be thought of as melts of {\it concentration blobs} so that the behavior on the lengthscales smaller than the blob size is similar to diluted solutions, and that on the larger lengthscales is similar to melts. Therefore, one can expect that a chain in semidilute solution needs to contain of order $N_e$ concentration blobs (as opposed to $N_e$ monomer units in a melt) in order for entanglements to form. Using \eq{radius_semidilute} for the number of monomers in the blob, one deduces that entanglements in semidilute solutions are formed for chains longer than $\sim N_e \phi^{-5/4}$ monomer units ($\phi$ here is the volume fraction of polymer in solution). 

This consideration allows us to formulate applicability limits of different theories. Dynamics of isolated chains in the abundance of solvent is correctly described by the Zimm model, except for very short chains and/or very viscous fluid, where non-draining is not yet achieved and some intermediate regime between Rouse and Zimm may occur. As the concentration of polymer in solution increases, the chains start to overlap, shifting from dilute to semi-dilute regime. In the semi-dilute regime volume interactions (including hydrodynamic interactions) get screened on large scales, and large scale dynamics of the chains in semi-dilute solutions is reasonably well described by the Rouse model until entanglements start to play role at larger concentrations. Thus, relaxation of small chain fragments, shorter than the size of the concentration blob $ s \leq \phi^{-5/4}$ is well-described by the Zimm model for swollen chains, the intermediate scales $N_e \phi^{-5/4} \leq s \leq \phi^{-5/4}$ relax according to the Rouse model, and at very long lengthscales  $s > N_e \phi^{-5/4}$ entanglements become important. Finally, in polymer melts, where $\phi = 1$, the Zimm regime disappears, relaxation of chains and chain fragments shorter than $N_e$ occurs according to the Rouse model, and the dynamics of longer chains is influenced by entanglements. 

To describe the dynamics of long linear chains a new approach allowing for the role of chin entanglements, is needed. The formulation of such an approach, know as reptation theory, is one of the great triumphs of the polymer physics of 1970s, achieved, for the most part, by P.-G.~de Gennes\cite{degennes_reptation}, S.F. Edwards, and M. Doi\cite{doi_edw1,doi_edw_2}. In the next subsection we give a very brief review of this theory. 

\begin{figure}[!t] \centering
\includegraphics[width=0.6\textwidth]{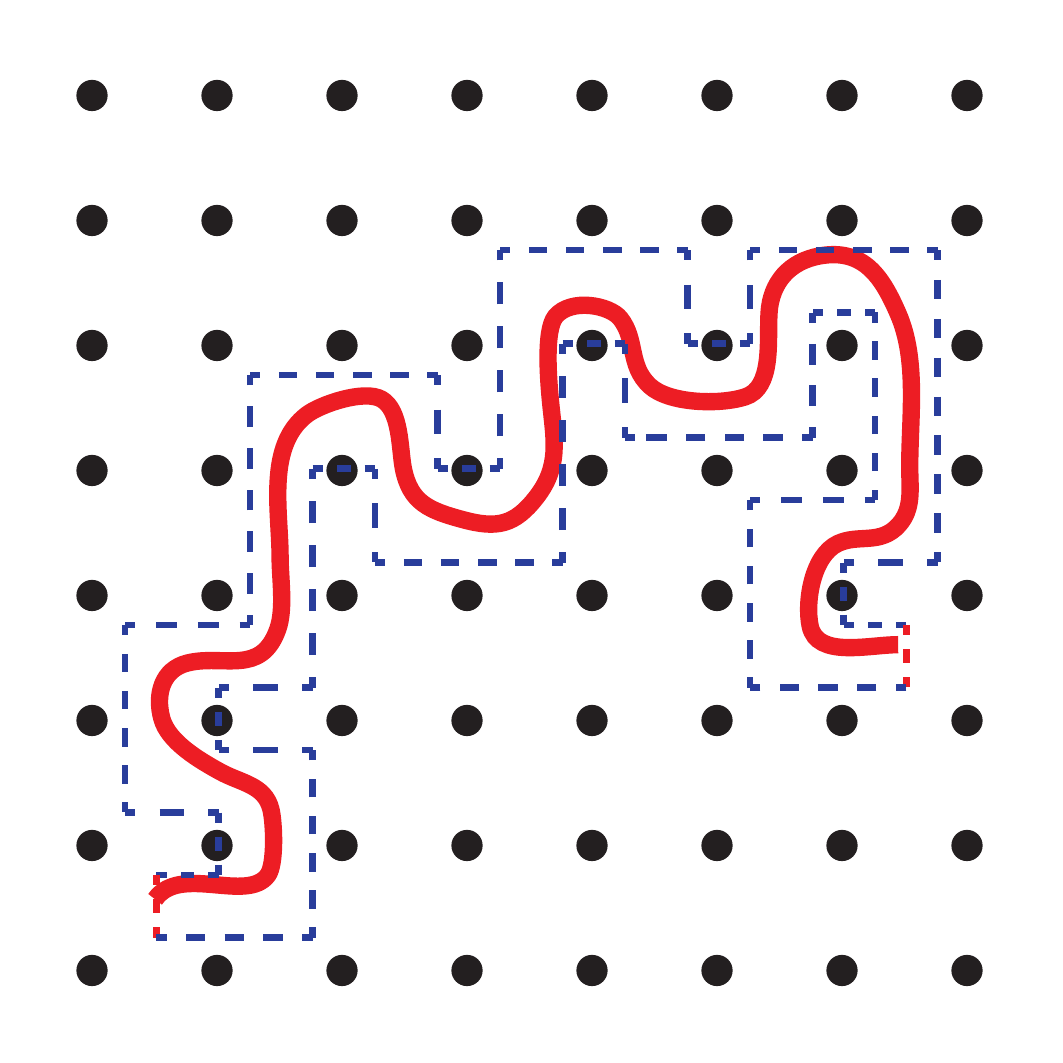}
\caption{Linear chain moving in the array of impenetrable obstacles. Obstacles form an effective tube, shown in blue dashed line, which the chain can leave only by diffusive movement of its ends, but not of its middle parts.}
\label{fig:6}
\end{figure}

\subsection{ Reptation theory}

To get some idea about chain movement in the presence of topological constraints let us once again return to the model problem of ``chain in array of obstacles'' discusses in section 1.6.1. In \fig{fig:6} we depict a linear chain in an array of obstacles. It is immediately clear\cite{degennes_reptation} that the chain movement is confined more or less to the tube-like region delineated by the dashed line. More precisely, while the chain can freely move within this region, the large-scale movement of the middle parts of the chain in the direction perpendicular to the tube is fully suppressed by the obstacles. Therefore the chain has to move by transferring its folds along the tube in a way akin of the movements of snakes, and diffusive movement of the chain ends through the end of the tube which are marked red in the \fig{fig:6}. Such a movement is known in the literature under the name of reptation, and the tube is called ``reptation tube''.

The dynamics of long chains in a polymer melt is very similar to that of a chain in an array of obstacles: entanglements with other chains form an array of unmovable obstacles which chain cannot pass, while away from these obstacles the movement is free and is locally well described by the Rouse model. One can think of a chain moving in the reptation tube as of a Rouse chain confined to a long winding cylindrical cavity. 

Note, however, that although entanglements are long-living, they are not permanent: they exist only up to the moment when the end of one of the constituting chains slips over it. The timescale at which this process, known as tube renewal, occurs is analogous to Rouse time in unentagled polymer systems: it is a maximal relaxation time of the chain, at times larger than tube renewal time the chain can move at distance larger than its original radius and the movement of monomer units is dominated by diffusion of the chain as a whole. In turn, on smaller timescales the entanglements play a role of effective cross-links capable of transmitting stress. As a result, polymer melts, while being viscous liquids with respect to low-frequency perturbations, behave as elastic solid objects when reacting to high frequency external force, the property known as viscoelasticity of polymer melts. 

To estimate the characteristic time of tube renewal $\tau^*$ one can proceed as follows. In order to renew the tube polymer should reptate diffusively out of it, therefore
\be
\tau^* \sim \frac{\Lambda^2}{D_{\text{tube}}}
\ee 
where $\Lambda$ is the length of the tube, and $D_{\text{tube}}$ is the diffusion coefficient for the movement of the chain along the tube. Now, within the confinement of the tube Rouse model still works, so the diffusion coefficient is still given by \eq{discm}, $D_{\text{tube}} \sim T \xi^{-1} N^{-1}$. The length of the tube can be estimated as follows. The tube consists of $N/N_e$ segments between entanglements, each of these segments (an ``entanglement blob'') takes the spatial size $a N_e^{1/2}$ since chains in the melt have ideal conformations according to the Flory theorem. Therefore,
\be
\Lambda \sim \frac{N}{N_e} aN_e^{1/2}
\ee
and
\be
\tau^* \sim \frac{Ta^2}{\xi} \frac{N^3}{N_e} \sim \tau_R \frac{N}{N_e}
\ee

Consider now the behavior of monomer displacement with time, $x^2 (t)$. In the reptation dynamics, there are four important timescales, instead of two in the Rouse model, and, subsequently, five important regimes with different behavior of $x^2(t)$. The important timescales are: the microscopic time $\tau_0$, given by \eq{microtime}; the maximal relaxation time of a chain fragment of length $N_e$, 
\be
\tau_{\max} (N_e) = \tau_0 N_e^2;
\ee
the Rouse time $\tau_R$; and the tube renewal time $\tau^*$ defined above. 

The time $\tau_{\max} (N_e)$ plays the crucial role because it is the timescale at which chain start feeling the presence of entanglements: before that the monomer displacements are too small to feel the fact that the chain is confined to the reptation tube. Therefore, on shorter timescales the entanglements are not yet felt and the dynamics is indistinguishable from the Rouse dynamics:
\be
\begin{array}{rll}
x^2(t) &\displaystyle \sim a^2 \frac{t}{\tau_0} &\text{ for } {t<\tau_0} \medskip \\
x^2(t) &\displaystyle \sim a^2 \left(\frac{t}{\tau_0}\right)^{1/2} &\text{ for } {\tau_0<t<\tau_{\max} (N_e)}
\end{array}
\label{reptation_short}
\ee
At larger times polymer is moving as a Rouse chain in confined geometry. If one introduces a curvilinear coordinate along the tube axis, the displacement along this coordinate would still be proportional to the square root of time, as in the second line of \eq{reptation_short}. But the reptation tube itself has a shape of random Brownian trajectory reflecting the fact that polymer chains in the melt take ideal conformations. Therefore, returning from this auxiliary curvilinear coordinate system to the original Cartesian frame of reference one gets
\be
\begin{array}{rll}
x^2(t) &\sim & \displaystyle a^2 N_e \left(\left( \frac{t}{\tau_{\max} (N_e)}\right)^{1/2}\right)^{1/2} \medskip \\
& \sim & \displaystyle a^2 N_e^{1/2} \left(\frac{t}{\tau_0}\right)^{1/4} \text{ for } {\tau_{\max} (N_e)<t<\tau_R}
\end{array}
\label{reptation_8}
\ee
For larger times all the Rouse modes of the chain have relaxed, but the tube is not renewed yet. Therefore the chain performs a normal diffusion along a winding tube, the mean-square displacement grows linearly with $t$ in the curvilinear coordinate frame, and proportional to $t^{1/2}$ in the external Cartesian frame:
\be
x^2(t) \sim a^2 \left( \frac{N_e}{N}\right)^{1/2} \left(\frac{t}{\tau_0}\right)^{1/2} \text{ for } {\tau_R<t<\tau^*}
\label{reptation_4}
\ee
Finally, at times longer than tube renewal time all the entanglements are relaxed, and the chain just simply performs normal diffusion 
\be
x^2(t) \sim a^2 \frac{N_e}{N^2} \frac{t}{\tau_0}
\label{reptation_2}
\ee
with effective diffusion coefficient proportional to $N_e N^{-2}$ instead of $N^{-1}$ in Rouse model. Note that prefactors in formulae \eq{reptation_8}, \eq{reptation_4}, and \eq{reptation_2} can be easily derived from the condition that different regimes should merge smoothly with each other.

\subsection{Entanglements in topological states of polymers}

In the light of recent developments in the understanding of the topological states of polymer systems, which we have reviewed in section 1.6, a natural question arises: what is the role of entanglements in the dynamics of these states? For the time being, this question is still open, with several competing theoretical approaches giving different answers, and with the results of computer simulations being rather inconclusive. Let us briefly overview the arguments of the different sides of the debate.

On the one hand, let us remind that the very existence of the topological states is due to the non-phantomness of polymer chains. If one allows chains to freely go through each other, the melt of nonconcatenated rings -- the archetypical topological polymer system -- will quickly relax to a state where the ring conformations will be ideal and all but indistinguishable from the conformations of linear chains in the melt. Therefore, it seems obvious that entanglements and topological interactions must play some important role in the chain dynamics as well.

On the other hand, one may argue that the role of topological interactions in systems like nonconcatenated rings actually is to force system into a state where there are as little entanglements as possible. The resulting conformations are compact, subchains form small compact domains (see \fig{fig:3}) and it is not clear how entanglements, at least in the traditional sense of the word, can actually form in such conformations\footnote{However, the fact that the compactized rings and other topological conformations of polymers have very developed surfaces additionally complicates the matter}. As a result one may argue that the role of entanglements in these states is actually just to act as a compressing external force which keeps the conformation of the chain compact, and it does not put any anisotropic constraints on the dynamics of the chain, just the constraint of preserving the constant fractal dimension of the packing.

As a result of the debate, several competing theories of the dynamics of topological states have been suggested recently. One group of authors, including ouselves\cite{tamm15, tamm17}, develop a Rouse-like model for the dynamics of crumpled states, exploiting the scaling ideas presented in section 2.2.1, and their analytical representation in terms of the beta model. Other authors\cite{smrek_dynamics, ge16} suggest two different versions of reptation-like theory to describe the dynamics of nonconcatenated rings. The competing theories predict the power-law behavior $x^2(t) \sim t^{\alpha}$ of the monomer displacement in the most interesting intermediate regime, but with very different values of $\alpha$: $\alpha = 0.4$ for the Rouse-like theory, and $\alpha \approx 0.26 - 0.28$ for the two versions of reptation-like theories. 

The situation is additionally complicated by the fact that both critical exponents of around 0.4 \cite{tamm15,jost_private} and around 0.28\cite{halverson11_2} have been reported in computer simulations, raising the suspicion that there is a possibility that there exist an additional governing parameter, akin to $N_e$ in the melts of linear chains, which separates regions of the Rouse-like and reptation-like dynamics. In particular, such a governing parameter might be especially relevant for the non-equilibrium topological systems like chromosomes in living cells (which in experiments tend to show dynamics with the exponent of 0.4 \cite{garini3, lagomarsino1}): immediately after cell division they are in very compact states which seemingly corresponds to Rouse-like dynamics, but with time going on the territories become somewhat more disperse which might at some point lead to a shift towards more reptation-like behavior. 

Anyway, the situation in the field is still very unclear. Note also, that large-$N$ asymptotic for the topological systems are reached only very slowly, which means that any comparison with experiments or simulations should be taken very carefully indeed. Hopefully, upcoming years will bring some clarity, and we will finally be able to fully understand both static and dynamical properties of the fascinating topological states of polymer chains. At the very least, we definitely have many exciting developments ahead.

\section*{Acknowledgements}

We are grateful to all the numerous colleagues with whom over the years we have discussed the topics covered in this chapter, most importantly to A.~Chertovich, M.~Cosentino-Lagomasrsino, I.~Erukhimovich, R.~Everaers, M.~Gherardi, A.~Grosberg, D.~Jost, E.~Kepten, A.~Khokhlov, R.~Metzler, L.~Mirny, A.~Mironov, L.~Nazarov, S.~Nechaev, G.~Oshanin, and Y.~Rabin. The more elementary parts of this chapter are largely based on a lecture course on polymer physics which M.V.T. is reading at Moscow State University for the last 10 years. It is a pleasure to thank all the students who attended the course over the years for their feedback which allowed to significantly improve the logic of presentation. We thank P.~Khalatur and L.~Nazarov for the permission to use their illustrations in this text, and T.~Nizkaya for critically reading the manuscript. It is our particular pleasure to thank Yu.~Holovatch for his invitation to write this chapter, and for his patient encouragement throughout the process of its preparation. This work was undertaken with financial support of the EU-Horizon2020 IRSES project DIONICOS (612707) and the Junior Leader grant of the BASIS foundation.

\end{document}